\documentclass[twocolumn,showpacs,showkeys,preprintnumbers,amsmath,amssymb,pra]{revtex4}

\usepackage{graphicx}% Include figure files
\usepackage{graphics}
\usepackage{epsfig}
\usepackage{dcolumn}% Align table columns on decimal point
\usepackage{bm}% bold math

\newcommand{\beq}{\begin{equation}}
\newcommand{\eeq}{\end{equation}}
\newcommand{\bqa}{\begin{eqnarray}}
\newcommand{\eqa}{\end{eqnarray}}
\newcommand{\nn}{\nonumber}
\newcommand{\nl}[1]{\nn \\ && {#1}\,}
\newcommand{\erf}[1]{Eq.~(\ref{#1})}

\newcommand{\ket}[1]{|{#1}\rangle}

\newcommand{\an}[1]{\left\langle{#1}\right\rangle}

\begin{document}
%\preprint{preprint number : }
\title{States for phase estimation in quantum interferometry}% Force line breaks with \\
%HERE
%==============================================================================
% Personal Information
%==============================================================================
\author{Joshua Combes}
\email{J.Combes@griffith.edu.au}
\author{H. M. Wiseman}%
\email{H.Wiseman@griffith.edu.au}

\affiliation{Centre for Quantum Computer Technology, Centre For
Quantum Dynamics, School of Science, Griffith University, Brisbane
4111, Queensland, Australia.}
\date{\today}

%==============================================================================
% Abstract
%==============================================================================
\begin{abstract}
Ramsey interferometry allows the estimation of the phase $\varphi$ of rotation
of the pseudospin vector of an ensemble of two-state quantum systems.   For
$\varphi$ small, the noise-to-signal ratio scales as the spin-squeezing
parameter $\xi$, with $\xi<1$ possible for an entangled ensemble. However
states with minimum $\xi$ are not optimal for single-shot measurements of an
arbitrary phase. We define a phase-squeezing parameter, $\zeta$, which is an
appropriate figure-of-merit for this case. We show that (unlike the states that
minimize $\xi$), the states that minimize $\zeta$ can be created by evolving an
unentangled state (coherent spin state)  by the well-known 2-axis
counter-twisting Hamiltonian. We analyse these and other states (for example
the maximally entangled state, analogous to the optical ``NOON" state
$|\psi\rangle = (|N,0\rangle+|0,N\rangle)/\sqrt{2}$) using several different
properties, including $\xi$, $\zeta$, the coefficients in the pseudo angular
momentum basis (in the three primary directions) and the angular Wigner
function $W(\theta,\phi)$. Finally we discuss the experimental options for
creating phase squeezed states and doing single-shot phase estimation.
\end{abstract}



                             % 42.50.Dv Non classical states

                             % 42.50.Lc Quantum fluctuations

                             % 07.60.Ly Interferometry

                             % 06.30.Ft time and frequency

                             % 39.30.+w spectroscopic techniques

%Use showkeys class option if keyword


\maketitle

%==============================================================================
% Begin Main Body
% Section I - The Introduction
%==============================================================================
\section{\label{Intro}Introduction}

A spin squeezed state (SSS) \cite{Kitagawa,Wineland1}
is a collective (entangled) state of many individual spin-systems
such that a parameter $\xi^2$ is less than unity. The value $\xi^2 = 1$
is known as the standard quantum limit, as it is the
value it would have if all of the individual spin vectors were
unentangled, and oriented in the same direction:
 a coherent spin state (CSS) \cite{Kitagawa,Wineland1}.

 Various authors have suggested that spin squeezed states could
improve the precision of various measuring devices
\cite{Wineland1,Kitagawa}. In particular, there is a reduction of quantum noise in
Ramsey interferometry  \cite{Wineland1} by a factor $\xi^2$,
as verified  experimentally by Meyer {\em et al} \cite{expMRKSIMW}.
 One may be inclined to think that the parameter $\xi^2$ is the only one that
 matters for spin squeezed states. Here we argue that
its generality has been over stressed.

To be specific, if we wish to use a state for a single-shot measurement of the
angle $\varphi$ of rotation of the state around some axis of the Bloch sphere,
and there is no prior information about $\varphi$, then the maximally
spin-squeezed state is not the best state. This point has been previously made
in Ref.~\cite{BerryWisemanBreslin}. Here we delve into this issue in more
detail, defining a new parameter, $\zeta^2$, which we call phase-squeezing. We
investigate this, $\xi^2$, and several other characteristics of the CSS and
five different entangled spin-states. Through this we elucidate the relation
between concepts such as spin squeezing, phase squeezing, NOON states, and
phase estimation.

The states that give optimal single-shot precision were identified in Ref.
\cite{BerryWiseman1}, where a practical near-optimum single-shot estimation
scheme was also proposed. A method for engineering these optimal states for
optical interferometry was also suggested in the same  reference, but it would
be extremely challenging to implement. In this paper we have investigated a
more realistic proposal, to produce near optimal states for single shot quantum
interferometry in an ensemble of spins using the well-known 2-axis
counter-twisting Hamiltonian (2ACT) \cite{Kitagawa} (also see section
\ref{2ACTstates}) . The results are quite encouraging for the experimental
realization of optimal phase-squeezed states, in contrast to the situation for
maximally spin-squeezed states.

The structure of this paper is as follows. In Sec.~II we review
phase estimation by interferometry, and the limitations of $\xi^2$
as a figure of merit for this purpose. This motivates our introduction
of a new parameter, $\zeta^2$.
In Sec.~III we review the states we wish to consider: coherent
spin states, Yurke states, NOON states, optimal
phase-squeezed states, 2ACT spin-squeezed states, and 2ACT
phase-squeezed states. The last of these is introduced here for
the first time. In Sec.~IV we quantitatively investigate various properties
for each of these states: $\xi^2$, $\zeta^2$, the canonical phase
probability distribution, the state coefficients in the angular momentum
basis (in the $x$, $y$, and $z$ directions), and the angular
Wigner function $W(\phi,\theta)$. We emphasize the trends that can
be seen across the various states, especially using the Wigner function.
We conclude in Sec.~V with a discussion of the experimental
options for creating phase squeezed states and doing near optimal
single-shot phase estimation.

%==============================================================================
% Section II - Spin Squeezing and Interferometry
%==============================================================================
\section{Spin Squeezing and Interferometry\label{Interferometry}}

The equivalence between Ramsey interferometry and Mach-Zehnder (MZ)
interferometry  has been discussed at length \cite{Wineland2}. In the latter
case a field is introduced to the input ports of a MZ interferometer, then the
interferometer transforms the change in phase in one arm ($\varphi$) to changes
in intensity of the field at the output ports. If the $N$ quanta entering the
input ports are entangled the minimum detectable phase change scales as $1/N$,
the Heisenberg limit \cite{HollandBurnett}. This an increase in the signal to
noise ratio (SNR) when compared to a non entangled states (e.g. all quanta
entering at one port) of order $\sqrt{N}$. The non-entangled scaling of
$1/\sqrt{N}$ is known as the standard quantum limit.

Ramsey interferometry also measures a phase
$\varphi$: the angle of rotation of all spins about some axis.
Spin squeezing, a form of entanglement, enables this
phase to be estimated better than the standard quantum limit \cite{Wineland1}.
In principle this could enable a better
atomic clock, where an initial state is allowed to evolve (rotate)
and then the phase is estimated to give a measure of the time
elapsed from preparation. We shall now revisit the logic that led
to the formulation of the spin squeezing parameter as treated in
\cite{Wineland2}.

\subsection{Spin Squeezing Parameter $\xi^2$}\label{spinparam}

Consider a ensemble of $N$ two level atoms, a spin $J$ system,
where $N=2J$. The angular momentum operators are $J_i = (1/2)\sum
_{k=1}^N\sigma_i^k$ where $\sigma_i\in\{x,y,z\}$. For a precision
measurement on this system we are interested in the sensitivity of
our chosen input state to rotation. Following Wineland {\em et al}
\cite{Wineland2}, let the mean collective spin of the system be in
the $\hat{\textbf{x}}$ direction, i.e. $\langle \vec J\rangle
=\hat{\textbf{x}}|\langle J_x\rangle|$. An interaction between a
field and the spin system takes place, causing a rotation in
the initial state about the $z$ axis, of $\varphi$.

For $\varphi$ small, its value may be estimate by measuring
$J_y$, since
\begin{equation}\label{assumption}
\langle J_y \rangle = |\langle \vec J\rangle|\sin{\varphi}.
\end{equation}
If $M \gg 1$ measurements of $J_y$ are taken, then from the central limit
theorem the uncertainty in determining $\varphi$ from the results is
$\Delta\varphi = \Delta J_y/(\sqrt{M}\partial\langle
J_y\rangle/\partial\varphi)$. Using Eq. (\ref{assumption}) the precision in
$\varphi$ is
\begin{equation}\label{precision}
  \Delta\varphi = \frac{\Delta J_y}{|\langle J
  \rangle|\cos\varphi\sqrt{M}}.
\end{equation}
By inspection the minima of Eq. (\ref{precision}) occur when $\cos\varphi$ is
maximized. That is, for $|\varphi|\ll 1$.

Naively one would expect a minimum uncertainty coherent state \beq
\label{cohpsi} |\psi\rangle_{\rm coh} = |J,J\rangle_{x} \eeq to be a good state
to use for phase estimation as all the spins have their mean spin vectors
aligned. Here the notation $\ket{J,\mu}_k$ indicates the egienstates of $J_k$
(where typically $k=x,y,z$) with eigenvalue $\mu$. The result that the CSS
gives us is the standard quantum limit (SQL). So a ratio of the uncertainty in
the state under examination to the SQL will be defined as the squeezing
parameter $\xi$. The uncertainty for the coherent state is $\Delta\varphi_{\rm
coh} = (J/2M)^{1/2}/J=1/\sqrt{NM}$. Using this SQL, the squeezing parameter
becomes
\begin{equation}\label{xir}
  \xi =
  |\Delta\varphi|/|\Delta\varphi|_{\rm coh}=\sqrt{2 J}\frac{\Delta
  J_y}{|\langle \vec J\rangle|}.
\end{equation}
A system with a parameter $\xi^2$ less than unity is spin squeezed and
necessarily entangled \cite{WangSanders}.

Although states with small $\xi^2$ are good for estimating a small $\varphi$
from multiple ($M\to\infty$) measurements, they are not necessarily optimal for
single-shot measurements, or for measurements where $\varphi$ is not known to
be small. Consider the case where $N$ is even so that $J=N/2$ is an integer.
Then a state which comes close to minimizing $\xi^2$ is the Yurke like state
\cite{yurke, Wineland2}, \beq \label{yurkpsi}  |\psi\rangle_{\rm yur} =
\frac{\sin\alpha}{\sqrt
2}|J,1\rangle_{y}+\cos\alpha|J,0\rangle_{y}+\frac{\sin\alpha}{\sqrt2}|J,-1\rangle_{y}.
\eeq The minimum $\xi^2 \sim \sqrt{2/N}$ is achieved as $\alpha \to 0$. In this
limit the state is invariant under a rotation of $\pi$ around the $z$ axis.
That is, in a single shot measurement it would be impossible to distinguish
between a rotation of $\varphi$ and one of $\varphi+\pi$. An even more extreme
example is the so-called NOON states \cite{Wineland3}, defined as \beq
\label{suppsi} |\psi\rangle_{\rm NOON} =
(|J,J\rangle_{z}+|J,-J\rangle_{z})/\sqrt{2}. \eeq These states are so-called
because in the field representation the state is described as a superposition
of Fock states: $(\ket{N,0} + \ket{0,N})/\sqrt{2}$. Like the Yurke states,
these states allow a measurement of phase with a sensitivity $O(\sqrt{N})$
times better than a coherent state \cite{Wineland3,Huelga}. However they are
are invariant under a $z$-rotation of $2\pi/N$. Thus, in a single shot
$\varphi$ would already have to be known to an accuracy of $O(1/N)$ for these
states to be useful at all.
%Note that $\xi^2$ is not defined for these states. *****put later

\subsection{Phase squeezing parameter $\zeta^2$}\label{phaseparam}

If we wish to estimate the phase shift $\varphi$ from a single measurement,
with no prior information about the phase, the quantity we wish to minimize is
the uncertainty in that single shot estimate, not the signal to noise ratio or
$\xi$. We consider minimizing the uncertainty in an {\em optimal} measurement,
which requires a generalized measurement \cite{SandersMilburnZhang}, (although
see Ref.~\cite{smbdtpegg1}). If arbitrary unitaries can be implemented (as in a
quantum computer) and projective measurements are possible, then such
generalized measurements can be done \cite{GardinerCiracZoller}. Even without
such power (effectively that of a quantum computer), a measurement that
performs almost as well as an optimal measurement (on any state) can be
achieved by {\em adaptive} projective measurements on single spins (or quanta)
\cite{BerryWiseman1, BerryWisemanBreslin}.

The optimal or canonical measurement scheme
\cite{SandersMilburn95,SandersMilburnZhang} involves projection of the state
onto the phase states
\begin{equation}\label{phasestates}
  |J,\phi\rangle= (2J + 1)^{-1/2}\sum
_{\mu=-J}^{\mu=J}e^{-i\mu\phi}|J,\mu\rangle_z.
\end{equation}
The probability operator measure (POM) for such phase measurements is
\begin{eqnarray}\label{POM}
\nonumber  {E}(\phi)d\phi &=& \frac{2J+1}{2\pi}|J\phi\rangle\langle
J\phi|d\phi\\
&=&
\frac{1}{2\pi}\sum_{\mu,\mu'=-J}^{\mu=J}e^{i(\mu-\mu')\phi}|J,\mu\rangle_z\langle
J,\mu'|d\phi .
\end{eqnarray}

 Provided there is no prior phase information, such measurements are optimal
for {\em all} states for which the arguments of the coefficients in the
$\ket{J,\mu}_z$ basis are linear in $\mu$ \cite{Holevo} (as, for example, in
the phase state $\ket{J,\phi}$). This includes all states that have been
considered for quantum interferometry. Assuming that the initial state
$\ket{\psi}$ is oriented in the $x$ direction, the POM (\ref{POM}) defines the
probability distribution for $\phi$, the best estimate for the phase shift
$\varphi$, via
\begin{equation}\label{probdist}
  P(\phi)d\phi=\langle\psi|e^{+i\varphi J_z}{E}(\phi)e^{-i\varphi J_z} |\psi\rangle d\phi.
\end{equation}

In the canonical measurement scheme it is sensible, for cyclic
variables, to define uncertainty is in terms of the sharpness
\cite{BerryWiseman1}
\bqa \label{sharpness}
S &=&  \langle e^{i(\varphi-\phi)}\rangle=  \int d\phi P(\phi)e^{i(\varphi-\phi)} \\
&=& \sum_{\mu= -J}^{\mu= J} {}_z\langle J,
\mu+1|\psi\rangle\langle\psi|J,\mu\rangle_z.
\end{eqnarray}
As long as  $\ket{\psi}$ is oriented in the $x$ direction,
 $S$ is real and positive, and it is always less than 1.
 Unlike the variance, the sharpness respects the periodicity of
 the phase $\phi$. A state with a $S$
close to 1 has a low phase uncertainty, and vice versa. For such states, the
variance of $P(\phi)$, $\int_{\phi_0}^{\phi_0+2\pi} P(\phi)\phi^2 d\phi$,
 is given by the approximate formula
\beq V \simeq 2(1-S), \eeq provided that $\phi_0$ is not near the peak of
$P(\phi)$.  (Another variance can be defined using $S$, the Holevo variance
\cite{Holevo} $S^{-2}-1$ but for our purposes it is simpler to take a quantity
linear in $S$.)

 Now that we have a measure for phase uncertainty, we can define a new figure of merit,
 appropriate to a single-shot phase estimate with no prior information:
 \begin{equation}\label{zeta}
  \zeta^2 = 4J(1-S).
\end{equation}
Similar measures have been considered before; see for example
\cite{HradilMyska}. It could also be asked, why not scale this equation with
respect to the coherent state i.e.
\begin{equation}\label{zetap}
    \zeta^2_{\rm rc} = (1-S)/(1-S_{\rm coh}).
\end{equation}
The reason Eq. (\ref{zeta}) was chosen over
Eq. (\ref{zetap}) is that it is a more elegant definition and with
small exceptions, discussed in section \ref{Props}, the two
expressions are equal. That is, $\zeta^2 \approx 1$ for coherent states.

%==============================================================================
% Section III  - The States
%==============================================================================
\section{States}
To obtain a better understanding of phase squeezing and the short comings of
spin squeezing a number of properties across six test states will be compared.
The test states are a coherent state (\ref{cohpsi}), a Yurke state
(\ref{yurkpsi}), a NOON state (\ref{suppsi}), plus three new states we define
below in this section: the optimal phase squeezed state,  the optimal 2ACT-spin
squeezed state, and the optimal 2ACT-phase squeezed state.  All of the states
considered the mean spin direction is in the $\hat{\textbf{x}}$ or
$(\theta,\phi) = (0,\pi)$, with the exception of the NOON state, and all except
this state and the coherent state are spin-squeezed states, with `squeezing' in
the $\hat{\textbf{y}}$ direction. The properties to be examined across all
states include the phase and spin squeezing parameters as a function of the
number of particles; the
 Wigner function; state coefficients; and the phase
distribution.

\subsection{Optimal Phase Squeezed}
The maximally phase squeezed states can be found analytically
\cite{BerryWiseman1}. These states, that minimize $\zeta^2$, are given by
\begin{equation} \label{optpsi}
|\psi\rangle_{\rm opt} = \frac{1}{\sqrt{J+1}}\sum_{\mu
  =0}^{2J}\sin\left[ \frac{(\mu +1)\pi}{2J+2}\right]|J,\mu \rangle_{z}.
\end{equation}

\subsection{2ACT Squeezed States}\label{2ACTstates}
So far we have not been concerned about how to create squeezed states. One of
the earliest suggestions was to start with a coherent state $|\psi_{\rm
coh}\rangle$ and to evolve according to the so-called
 two-axis
countertwisting (2ACT) [1] Hamiltonian \begin{equation} H_{\rm 2ACT} = \hbar
\gamma (J_yJ_z+J_zJ_y)
\end{equation}  Here $\gamma$ is the strength of the interaction (A generalized Hamiltonian for spin squeezing has been
devised by Wang and Sanders [7]). This generates the unitary evolution operator
\begin{equation}\label{2acth}
U(\nu)  =\exp \left[ \nu\left( J_+^2-J_-^2\right)/8 \right].
\end{equation}
Here $\nu=4\gamma t$  is a scaled time (we use $\nu$ rather than $\mu$ as in
[1] to avoid confusion with the $|J,\mu\rangle$ basis states), and $J_\pm = J_y
\pm iJ_z$ are the raising and lowering operators in the
$\hat{\textbf{x}}$-direction. This Hamiltonian produces squeezing in the
$\hat{\textbf{y}}$-direction, and the mean spin remains aligned along the
$\hat{\textbf{x}}$ axis.

\begin{figure}[!h]
\epsfig{file=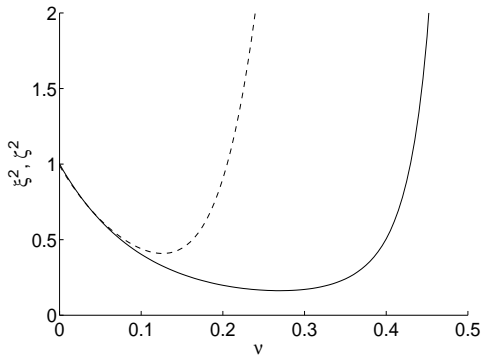,height=5cm,width=7cm} \caption{The parameters
$\xi^2$ (solid line), $\zeta^2$ (dashed line) are plotted for the state
$|J,-J\rangle$ evolving in time ($\nu=4\gamma t$) under the two axis counter
twisting Hamiltonian. $N= 20$.\label{fig1}}
\end{figure}

As is well known \cite{Kitagawa}, the state \beq \ket{\psi({\nu})}=
U(\nu)\ket{\psi_{\rm coh}} \eeq has a minimum in $\xi$ for an optimal value
$\nu=\nu_{\rm ss}$. This is shown in Fig.~\ref{fig1}. We denote the state for
this  value as
\begin{equation}
\label{ssspsi} \ket{\psi}_{\rm sss} = \ket{\psi(\nu_{\rm ss})}
\end{equation}
 and call it the 2ACT spin-squeezed state. As we show here (for the first
time), this phenomenon also occurs for $\zeta^2$. That is, $\zeta$ is minimized
for an optimal value $\nu=\nu_{\rm ps}$. We denote the state for this value as
\beq \ket{\psi}_{\rm pss} = \ket{\psi(\nu_{\rm ps})} \eeq and call it the 2ACT
phase-squeezed state.

\begin{figure}
\begin{center}
\epsfig{file=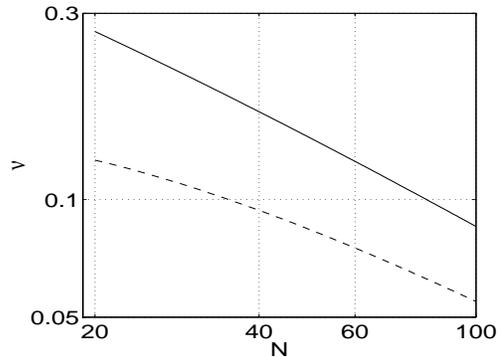,height=5cm,width=7cm} \caption{A log log plot of the
time ($\nu$) at which the maximal squeezing occurs for phase (dashed line) and
spin squeezing (solid line) parameters evolving under the 2ACT.\label{fig2}}
\end{center}
\end{figure}

As shown in Fig.~\ref{fig1}, the optimal value $\nu_{\rm ps}$ for phase
squeezing is less than that for spin-squeezing, $\nu_{\rm ss}$.  We have
numerically determined the optimal times $\nu$ for these two types of squeezing
and plotted the result in Fig.~\ref{fig2}. It has been previously shown that
$\nu_{\rm ss}$ scales as $\log_2(N)/N$ \cite{AndreLukin} in \cite{SGADM}. We
find specifically that $\nu_{\rm ss} \approx 1.25\log_2(N)/N$. As
Fig.~\ref{fig2} shows, $\nu_{\rm ps}$ appears to remain smaller than $\nu_{\rm
ss}$ even for large $N$, although the scaling law for the former is not known.

As is well known, and as we will show in the next section,
the 2ACT spin-squeezed states do not achieve the minimum
$\xi^2$ for a fixed $N$, and are not close to the states that do.
By contrast, we have found that the 2ACT phase-squeezed states
are almost identical to the optimal phase squeezed states
defined in \erf{optpsi}. As we will show in the next section,
the minimum $\zeta^2$ from the 2ACT-generated states is
practically indistinguishable from the minimum possible $\zeta^2$.

%==============================================================================
% Section IV  - Properties
%==============================================================================
\section{Properties\label{Props}}

\subsection{Spin Squeezing}\label{maxsqueezing}

Before discussing $\xi^2$ for our various states, we review the simple
proof in Ref.~\cite{Wineland2} for a lower bound.
Starting with the
uncertainty relation
\beq
\Delta J_x \Delta J_y \geq \an{J_z}^2/4
\eeq
and noting that $\langle J_y^2\rangle \leq J^2$, we get
\begin{equation}
\nonumber
  \langle J_x^2\rangle\geq\frac{\langle J_z\rangle^2}{4J^2}.
\end{equation}
Now substituting  this into \erf{xir} gives
\begin{equation}\label{scale}
  \xi^2\geq \left(\frac{\langle
  J_z\rangle^2}{4J^2}\right)\frac{N}{\langle
J_z\rangle^2}\geq \frac{1}{N},
\end{equation}
which is sometimes called the Heisenberg limit (HL). There is of course no
upper limit on $\xi^2$. Note that S\o rensen and M\o lmer \cite{SorensenMolmer}
produced a definitive paper on maximal squeezing for a state with given
$\an{J_x}$. The absolute minimum is obtained as $\langle J_x \rangle\rightarrow
0$, essentially reproducing the result of Yurke \cite{yurke}, which is \beq
\xi^2 =(1+N/2)^{-1} /\cos^2 \alpha \to 2/N. \eeq where the limit is taking
$\alpha \to 0$ and $N\to \infty$.

\begin{figure}[!h]
\begin{center}
\epsfig{file=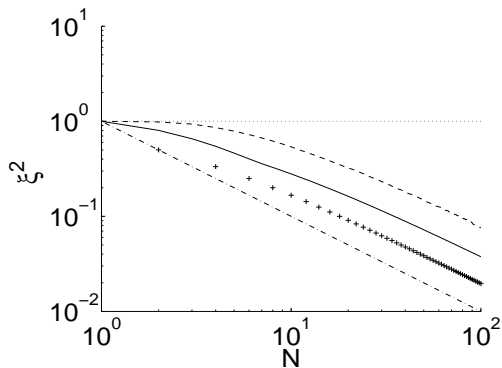,height=5cm,width=7cm} \caption{The parameter $\xi^2$,
plotted for our test states: dotted line for $\ket{\psi_{\rm coh}}$, dashed
line for $\ket{\psi_{\rm opt}}$, solid line for $\ket{\psi_{\rm sss}}$, and
$+$s for $\ket{\psi_{\rm yur}}$ (plotted only for $N$ even). The dash-dotted
line is $1/N$.  \label{fig3}}
\end{center}
\end{figure}

A plot of $\xi^2$ versus $N$ for our test states is shown in Fig.~\ref{fig3}.
Recall that for a coherent state $\xi^2=1$. For all the other states,
$\xi^2$ scales as $N^{-1}$ for large $N$, but with different coefficients.
The worst are the optimal phase squeezed states  (which as mentioned above are
practically identical to the 2ACT phase squeezed states), for which
$\xi^2 \approx 10/N$ for large $N$. Note also that for $N < 5$
the optimal phase-squeezed state is not significantly spin squeezed at all. %HERE
The NOON state is not included
as $\xi$ is undefined for this state.

\subsection{Phase Squeezing}

The absolute limit to phase squeezing is easy to analytically
obtain as a function of $N$. The maximally phase squeezed states
$\ket{\psi_{\rm opt}}$ have a sharpness given by \cite{BerryWisemanBreslin}
\beq\label{mpspd}
 S = \cos\left( \frac{\pi}{N+2}\right).
\eeq
For large $N$ this gives a phase-squeezing parameter
\begin{equation}\label{maxpslj}
\zeta^{2}\rightarrow \frac{\pi^2}{N}.
\end{equation}

As a comparison, the case for coherent states will be presented.
In the large $\langle J_x\rangle$ regime, $J_y/J\approx\phi$ and
$\langle J_y\rangle = 0$. For coherent states $\Delta J_x = \Delta J_y
= \sqrt{{J}/{2}}$ so we obtain
\begin{eqnarray}\label{exp}
\nonumber|\langle e^{i\phi} \rangle| &\approx& |\langle
e^{i{J_y}/{J}}
\rangle|\\
\nonumber &\approx& \langle 1 + i\frac{J_y}{J}-\frac{1}{2}
\frac{J_y^2}{J^2} \rangle\\
 &\approx&  1 -1/4J.
\end{eqnarray}
The phase squeezing parameter in the large $N$ limit is thus
\begin{equation}\label{pscoh}
\zeta^{2}\rightarrow 4J(1-(1 -1/4J)=1.
\end{equation}

\begin{figure}[!h]
\begin{center}
\epsfig{file=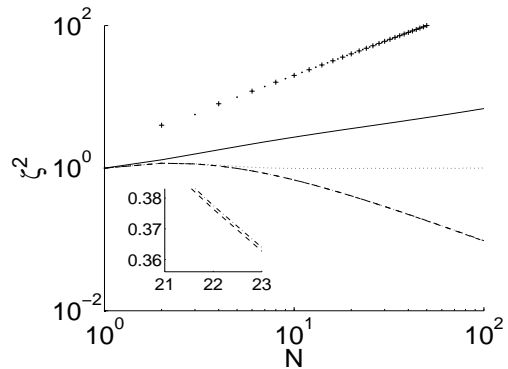,height=5cm,width=7cm} \caption{The parameter $\zeta^2$,
plotted for our test states: dotted line for $\ket{\psi_{\rm coh}}$, dashed
line for $\ket{\psi_{\rm opt}}$, dash-dotted line for $\ket{\psi_{\rm pss}}$,
solid line for $\ket{\psi_{\rm sss}}$, $\bullet$s for $\ket{\psi_{\rm NOON}}$
and $+$s for $\ket{\psi_{\rm yur}}$ (plotted only for $N$ even).
 The inset plot is
a close up of the same figure showing how close
$\ket{\psi_{\rm pss}}$ is to $\ket{\psi_{\rm opt}}$. \label{fig4}}
\end{center}
\end{figure}

A plot of $\zeta^2$ versus $N$ for our test states is shown in Fig.~\ref{fig4}.
Note that unlike $\xi^2$, $\zeta^2$ for  the coherent state does not exactly
equal one ---  It is noticeably larger than one for states with less than seven
particles. For large $N$ it asymptotes to 1 (from above) as expected from the
above analysis. From the linearity of \erf{zeta} in the state $\rho$, it
follows that $\zeta^2 > 1$ for all mixtures of coherent states. Thus $\zeta^2 <
1$ indicates phase squeezing (i.e. better than the standard quantum limit) and
hence entanglement (at least for states of well-defined $J$). This is similar
to the way $\xi^2 < 1$ indicates spin squeezing and hence entanglement,
although that has been shown even for states without well-defined $J$
\cite{Sorensen1}. However, unlike $\xi^2$, $\zeta^2$ does not drop below 1 even
for the optimal state until $N > 5$. The large $N$ scaling of $1/N$ is evident
for the optimal state, and the 2ACT phase squeezed state gives almost identical
results. But in contrast to the $\xi^2$ calculation, in this case all other
entangled states actually have $\zeta^2$ {\em increasing} with $N$. For the
NOON state $\zeta^2 = 2N$ and for the Yurke state this is very nearly true.
This is because of the
  the symmetry (or near symmetry) of these state implies that the
sharpness $S$ is zero for the NOON state and approaches zero
for the Yurke state. The result for the 2ACT-SSS will be
explained in the next subsection. %HERE

\subsection{Phase distribution}

Recall that the phase squeezing parameter $\zeta^2$ is a measure of the spread
of the phase distribution $P(\phi)$ of an optimal measurement of the phase
shift. To understand the results obtained above for $\xi^2$ we have plotted
$P(\phi)$ as the second row of Fig.~\ref{fig0} with $N=20$. In this figure we
have included all our states except $\ket{\psi_{\rm opt}}$ as it almost
identical to $\ket{\psi_{\rm pss}}$.

The coherent state does not contain any
surprises --- it is quite broad, corresponding to the standard quantum limit.
The next state, $\ket{\psi_{\rm pss}}$ has a much narrower peak as expected.
The third state, $\ket{\psi_{\rm sss}}$ has an even narrower central peak but
it also has significant side lobes and wings. It is these wings that have such
a deleterious effect on the performance of this state in a phase measurement,
with $\zeta^2$ larger than that for a coherent state. Unlike the first three
phase distributions, that of the Yurke state $\ket{\psi_{\rm yur}}$ is bimodal.
This shows clearly that this state can only determine $\varphi$ modulo $\pi$.
Even taking this into account, the side lobes in this distribution (like that
of the spin-squeezed state $\ket{\psi_{\rm sss}}$) also make this state
inferior to the coherent state for single-shot phase estimation, as demonstrated
in Ref.~\cite{BerryWisemanBreslin}. Finally, the
NOON state has $N$ peaks, and is the worst state of all in this context,
allowing only an estimate of the phase modulo $2\pi/N$. %HERE

\subsection{State Coefficients}
Again in Fig.~\ref{fig0}, the bottom three rows are the state
coefficients ${}_x\langle \mu|\psi\rangle$, ${}_y\langle \mu|\psi\rangle$, and
${}_y\langle \mu|\psi\rangle$ respectively. In each case the state has been
multiplied by an overall phase factor to ensure that the coefficients are positive.
It is instructive to examine the trends (across the states) in the coefficients
in the three cardinal directions separately.

For the $x$ coefficients, the coherent state has a single non-zero coefficient,
at $\mu = -J$.
Moving across, the phase squeezed state develops other non-zero
coefficients, and this is further developed in the spin-squeezed states,
where the non-zero coefficients stretch almost to $\mu=+J$. In the Yurke state
this goes even further, with the coefficients being bimodal,
with peaks at $\mu = \pm J$. The NOON state does not fit obviously into
this trend, as now there is a single peak at $\mu = 0$. In all of these states
the coefficients are zero for $\mu$ odd.

For the $y$ coefficients, the trend is even clearer from the coherent state
to the Yurke state: an initial symmetric Gaussian-like distribution becomes narrower and
narrower until it reaches a single non-zero coefficient at $\mu=0$.
Again, the NOON state appears anomalous, being a broad distribution.
Note however that unlike those for the other states, these coefficients are
zero for $\mu$ odd. In fact, these coefficients are identical to those for the $x$
direction, because the NOON state has no preferred phase.

It is only for the $z$ coefficients that a single trend appears to explain
the distribution for all states. To begin, the coherent state has a
symmetric Gaussian-like distribution (the same as that for its $y$ coefficients).
In opposition to the case of the $y$ coefficients, as the state becomes
more phase squeezed, this distribution becomes broader. This is
the expected phenomenon of antisqueezing. For the phase squeezed state
the distribution is sinusoidal [see \erf{optpsi}]. For the spin squeezed state
it becomes almost flat. Note however that at the ends of the distribution we see
the beginning of new trends: the even coefficients are larger than the odd ones,
and the largest coefficients are at $\mu = \pm J$. This trend is amplified in
the Yurke state, where all odd coefficients are zero, and the curve for the even
coefficients is concave up.
Finally, in the NOON state it is taken to the extreme where only the
$\mu = \pm J$ coefficients are non-zero.

\subsection{Wigner function}
\begin{widetext}

\begin{figure}
\begin{center}

%=====================================
% Label Row
%=====================================
\begin{minipage}{0.05\textwidth}
~~~~
\end{minipage}
\begin{minipage}{0.18\textwidth}
  Coherent
\end{minipage}
\begin{minipage}{0.18\textwidth}
Phase Squeezed
 \end{minipage}
\begin{minipage}{0.18\textwidth}
Spin Squeezed
 \end{minipage}
\begin{minipage}{0.18\textwidth}
Yurke
 \end{minipage}
\begin{minipage}{0.18\textwidth}
  NOON
\end{minipage}

%=====================================
% Wigner Row
%=====================================
\begin{minipage}{0.06\textwidth}
$W(\theta,\phi)$
\end{minipage}
\begin{minipage}{0.182\textwidth}
  \includegraphics[width=\textwidth, height=\textwidth ]{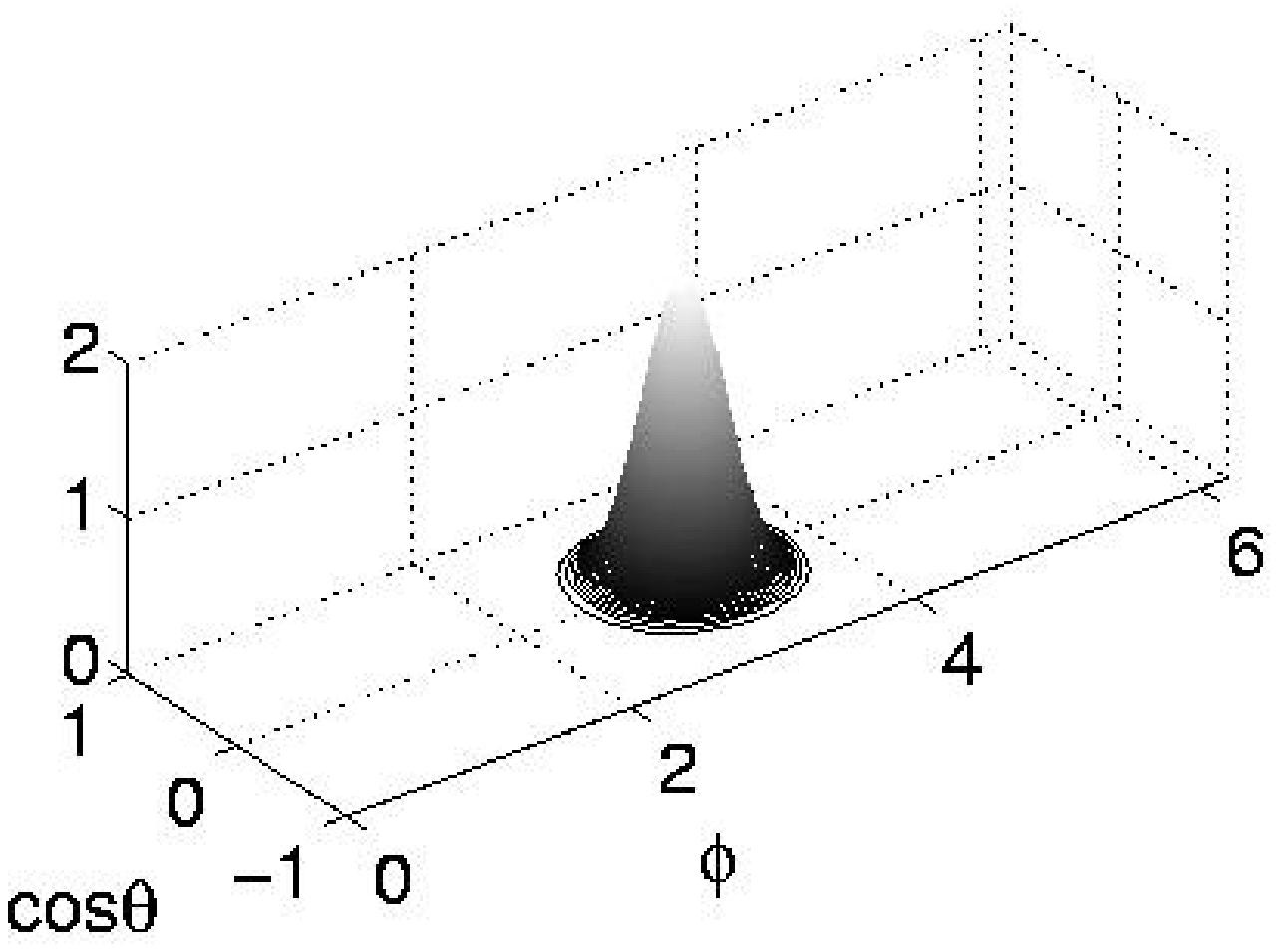}
\end{minipage}
\begin{minipage}{0.182\textwidth}
  \includegraphics[width=\textwidth, height=\textwidth ]{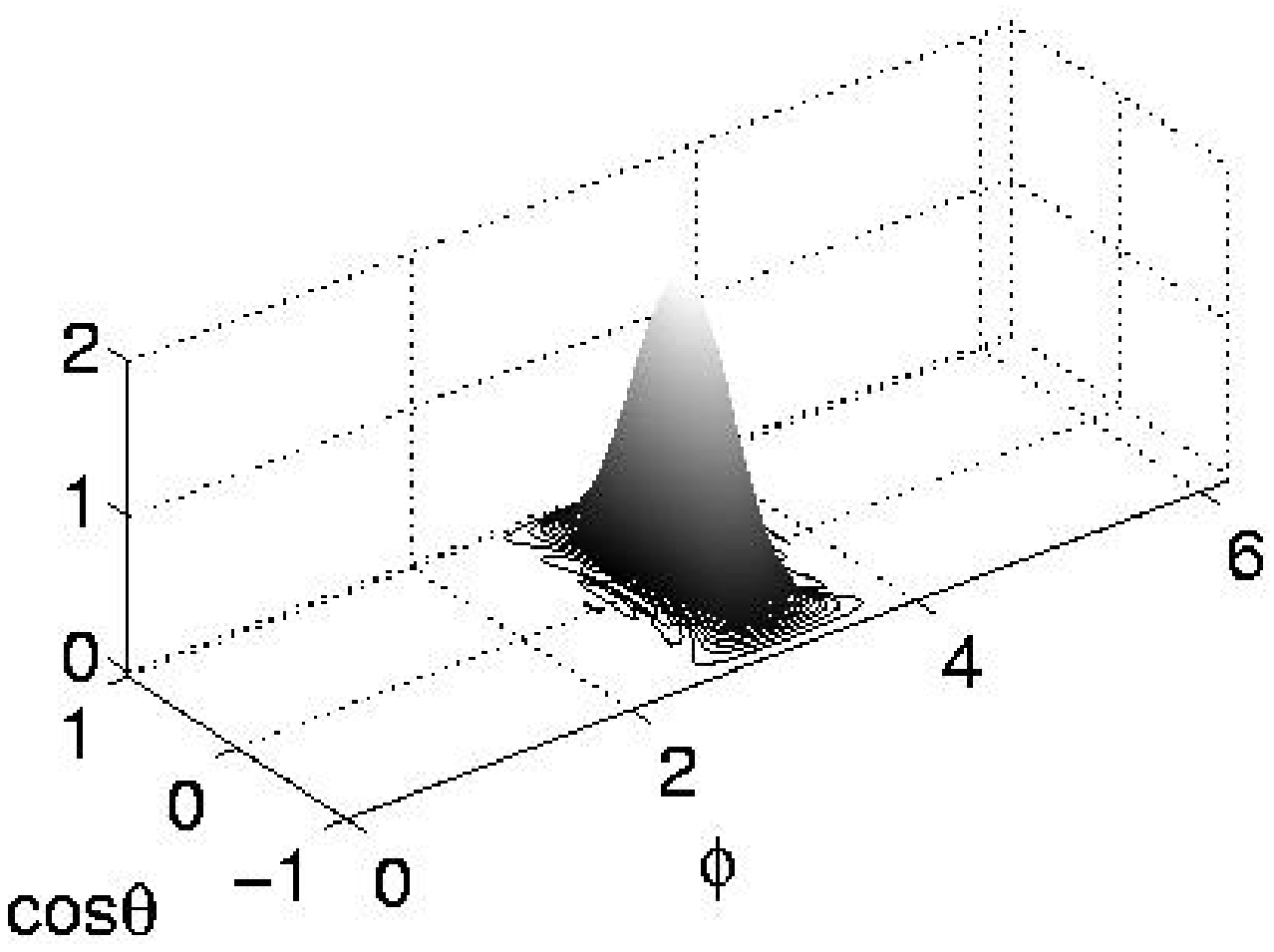}
 \end{minipage}
\begin{minipage}{0.182\textwidth}
  \includegraphics[width=\textwidth, height=\textwidth ]{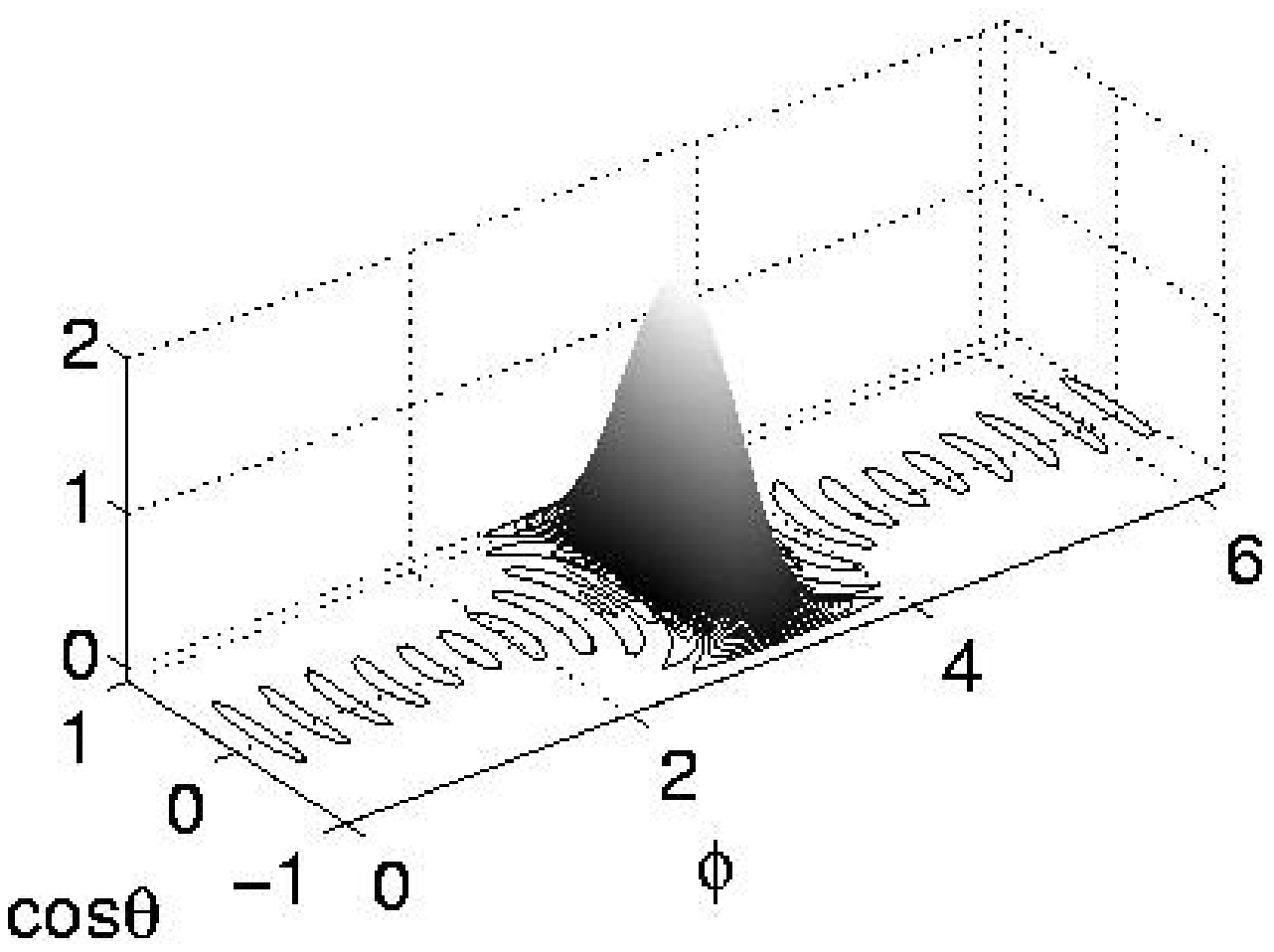}
 \end{minipage}
\begin{minipage}{0.182\textwidth}
  \includegraphics[width=\textwidth, height=\textwidth ]{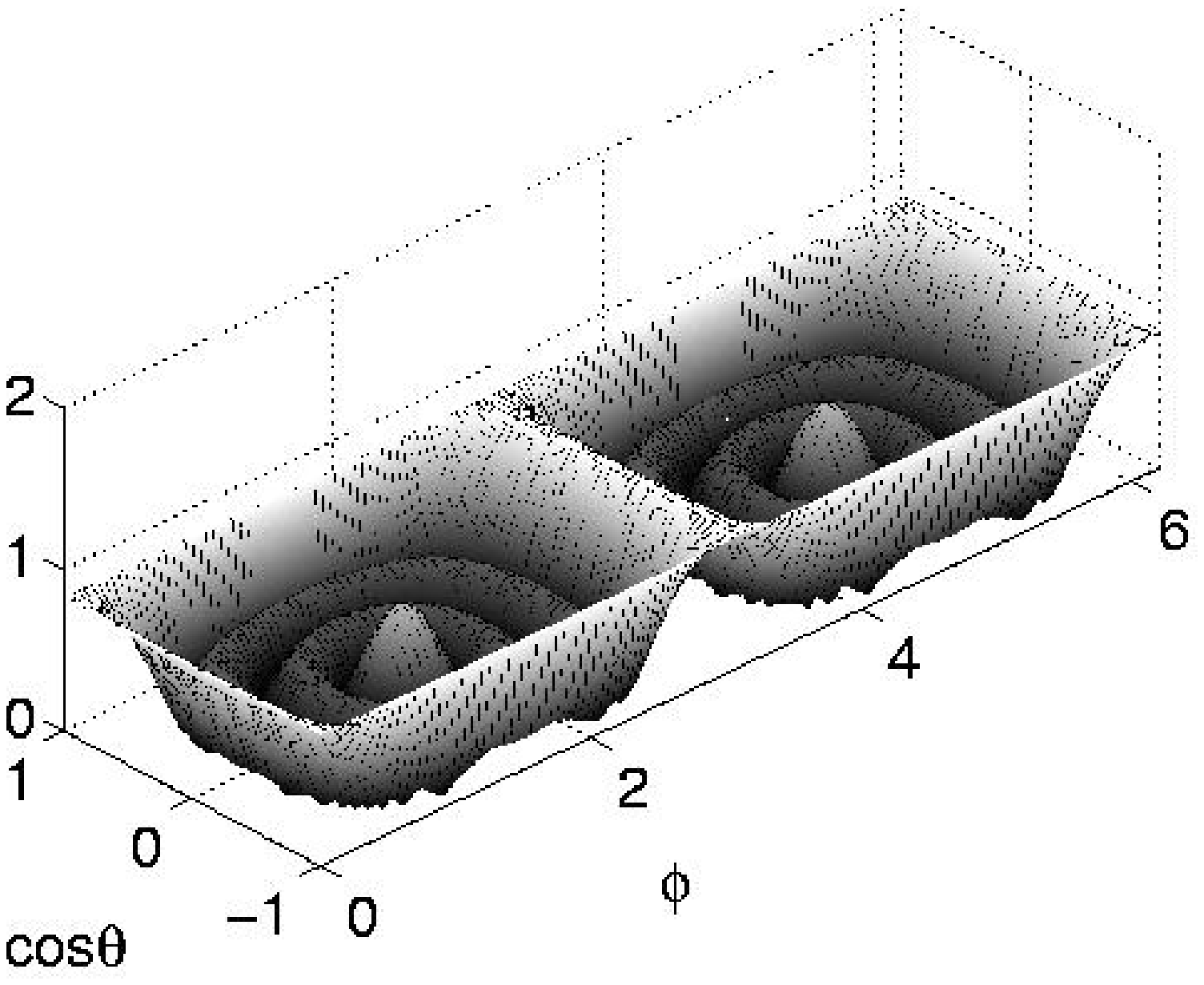}
 \end{minipage}
 \begin{minipage}{0.182\textwidth}
\includegraphics[width=\textwidth, height=\textwidth ]{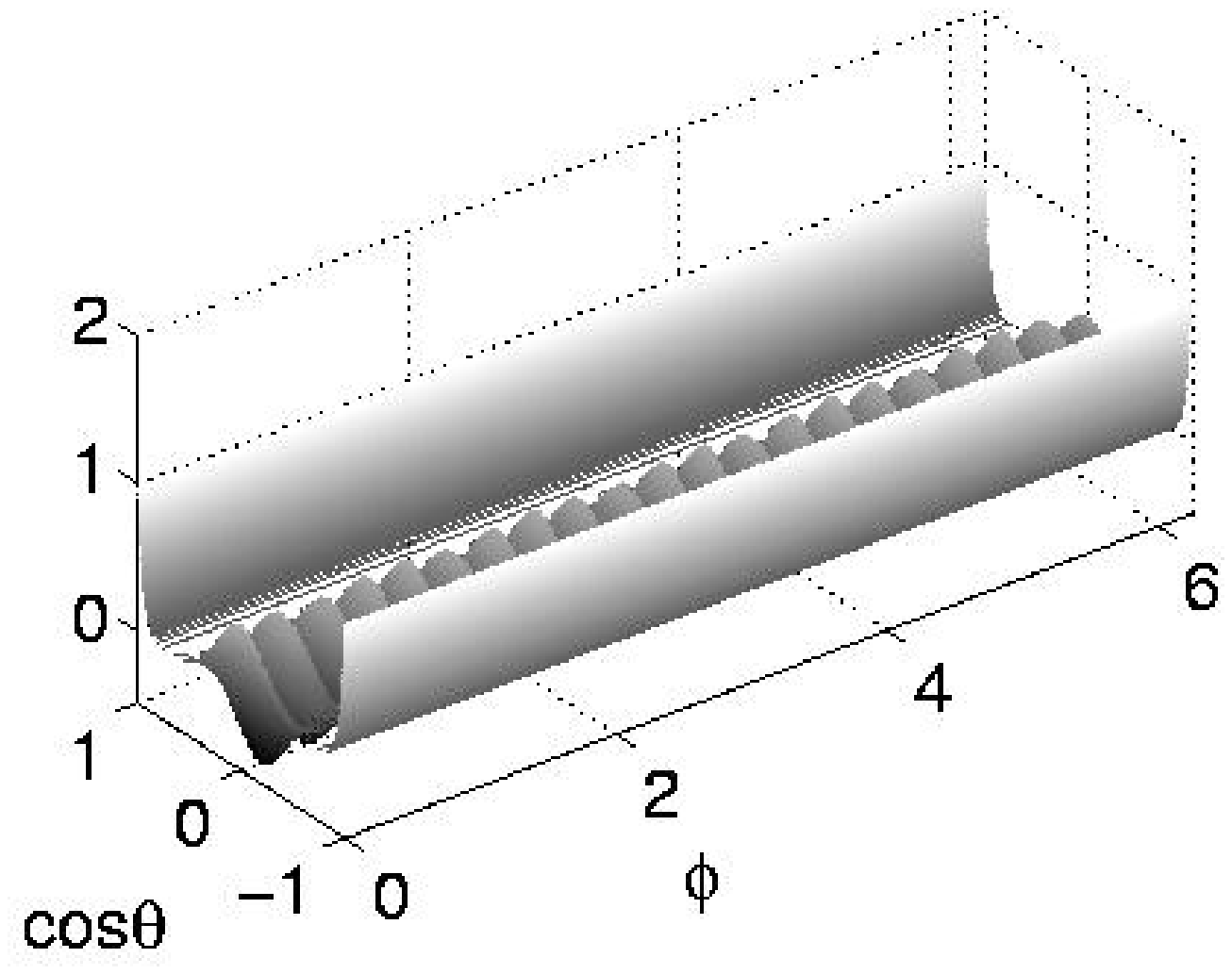}
\end{minipage}
%=====================================
% Phase Row
%=====================================
\begin{minipage}{0.04\textwidth}
$P(\phi)$
\end{minipage}
\begin{minipage}{0.182\textwidth}
  \includegraphics[width=\textwidth, height=\textwidth ]{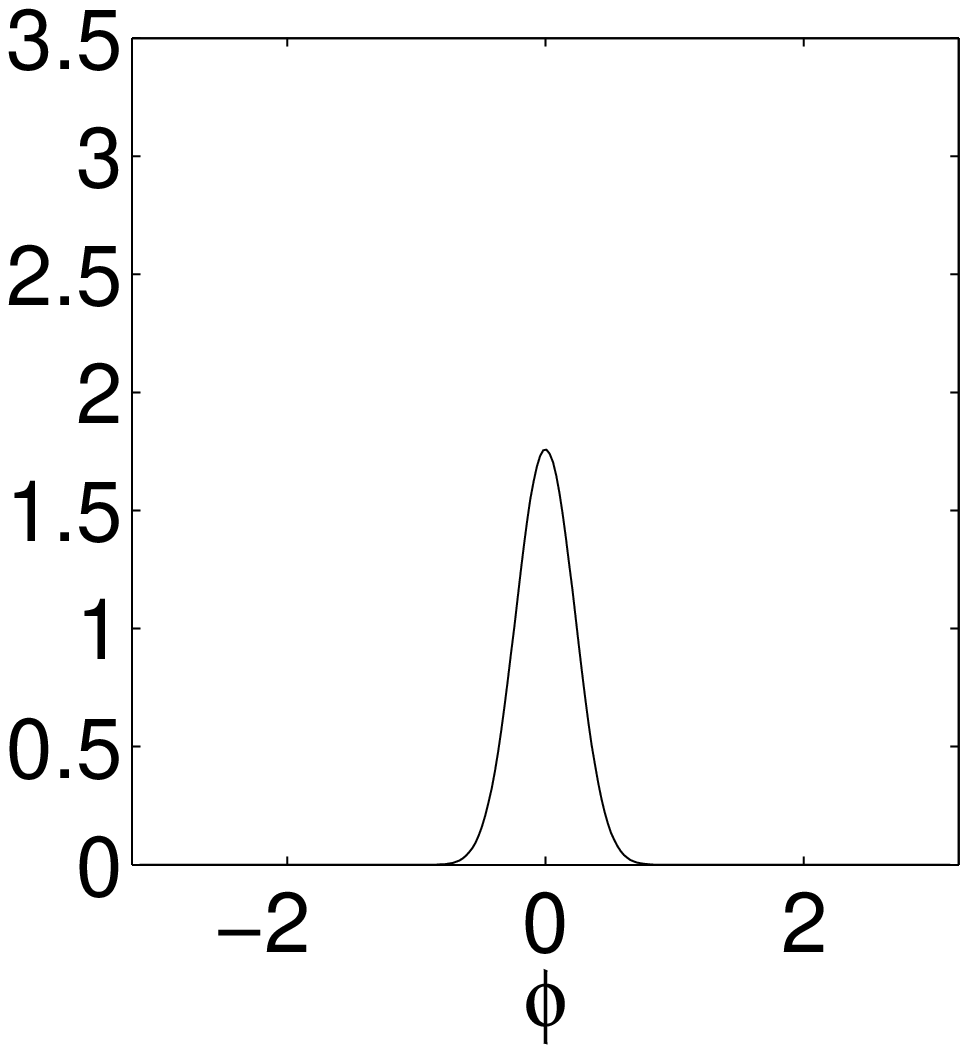}
\end{minipage}
\begin{minipage}{0.182\textwidth}
  \includegraphics[width=\textwidth, height=\textwidth ]{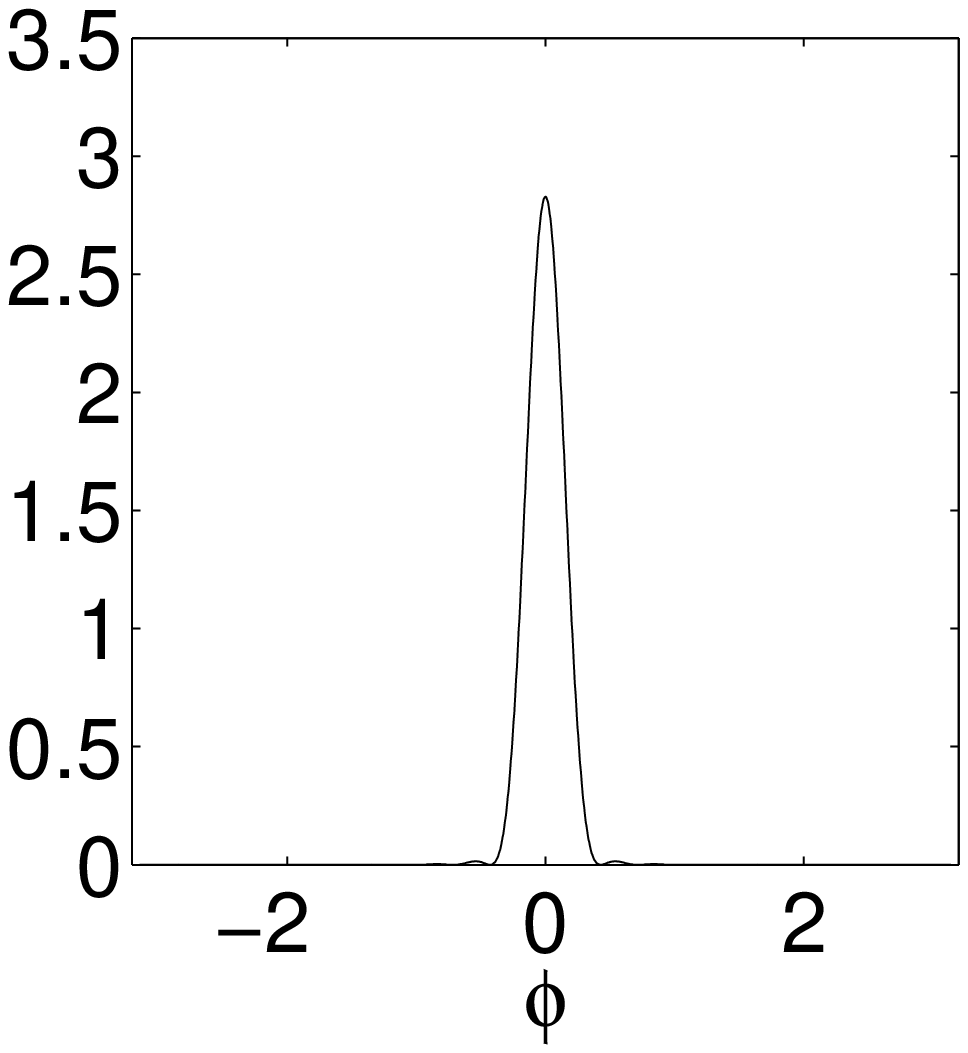}
 \end{minipage}
\begin{minipage}{0.182\textwidth}
  \includegraphics[width=\textwidth, height=\textwidth ]{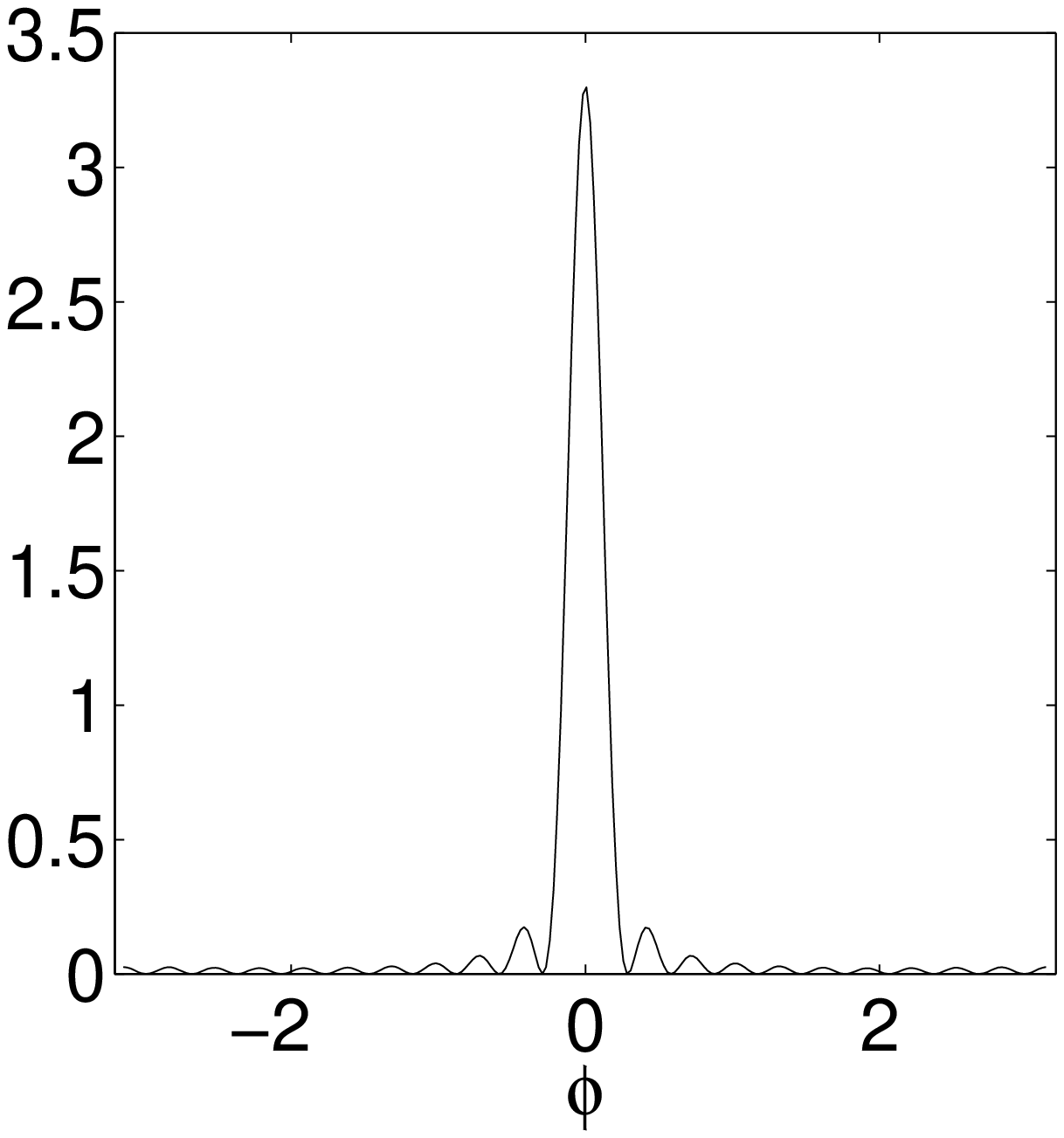}
 \end{minipage}
\begin{minipage}{0.182\textwidth}
  \includegraphics[width=\textwidth, height=\textwidth ]{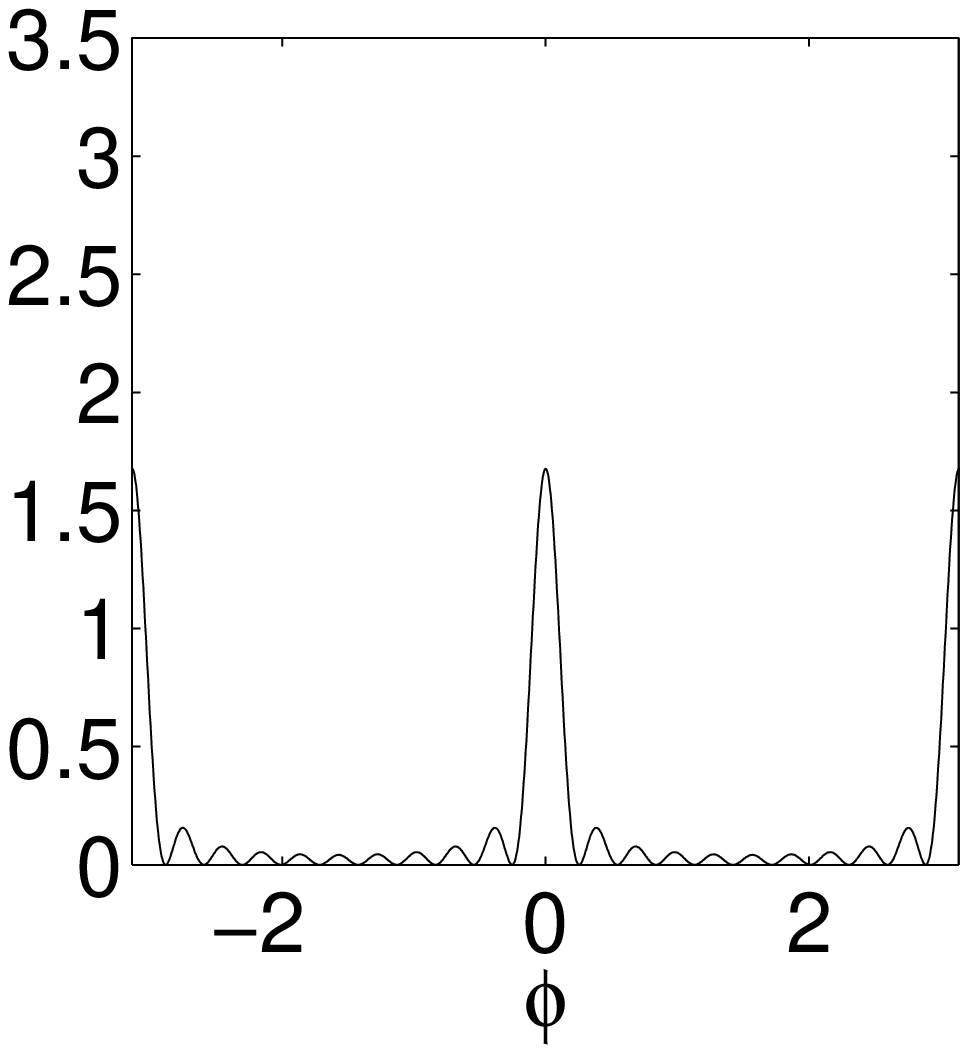}
 \end{minipage}
 \begin{minipage}{0.182\textwidth}
  \includegraphics[width=\textwidth, height=\textwidth ]{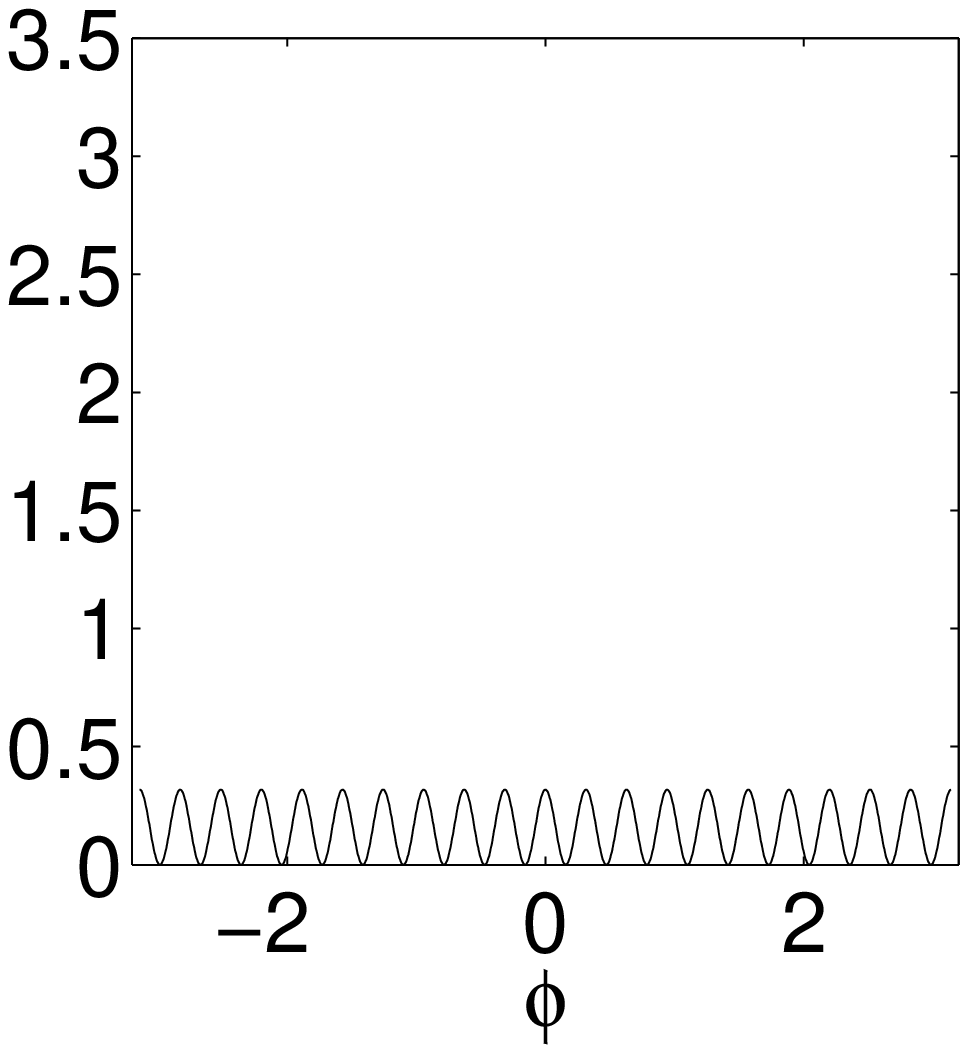}
\end{minipage}
%=====================================
% X Row
%=====================================
\begin{minipage}{0.04\textwidth}
$_x\langle\mu|\psi\rangle$
\end{minipage}
\begin{minipage}{0.182\textwidth}
  \includegraphics[width=\textwidth, height=\textwidth ]{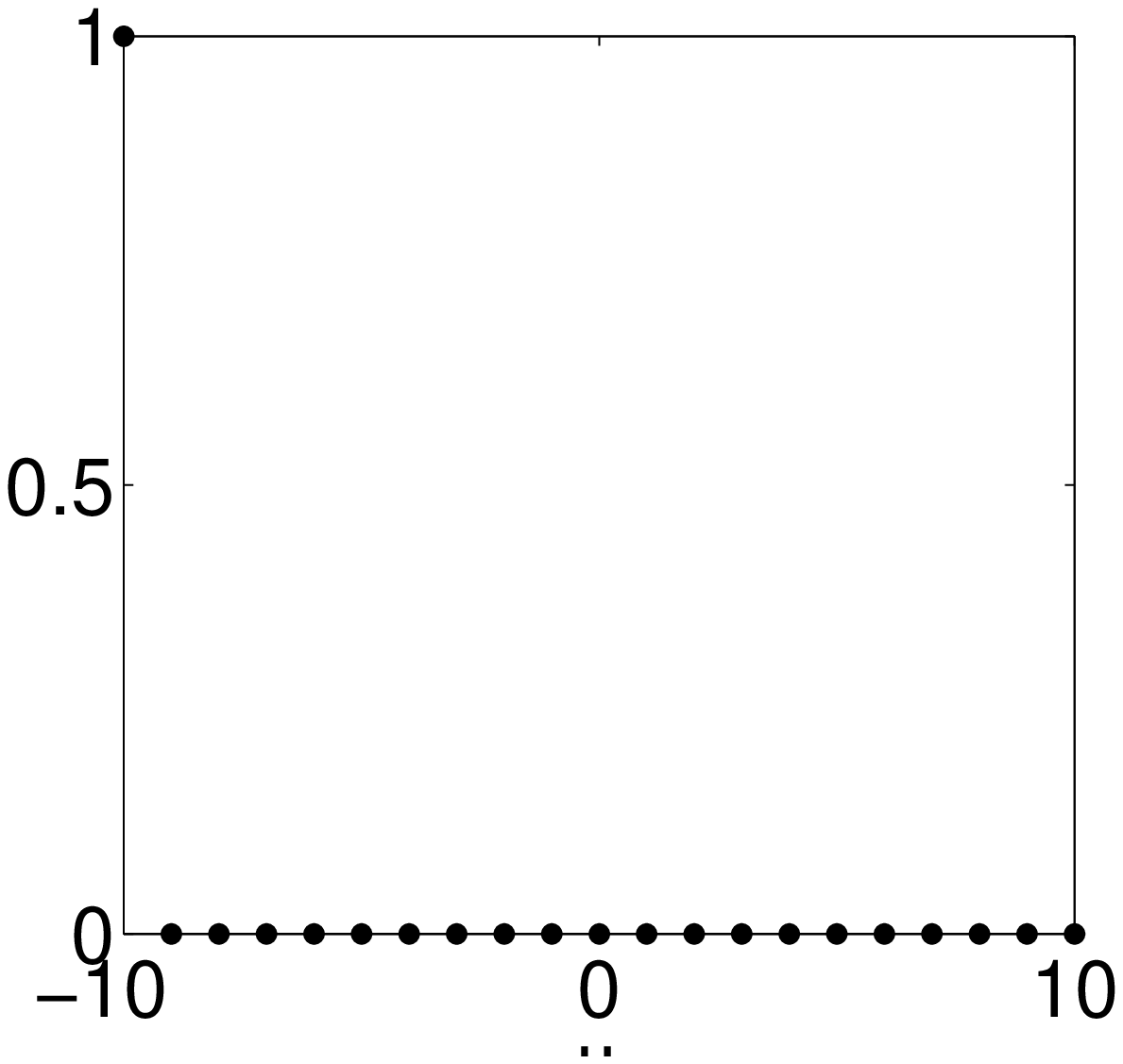}
\end{minipage}
\begin{minipage}{0.182\textwidth}
  \includegraphics[width=\textwidth, height=\textwidth ]{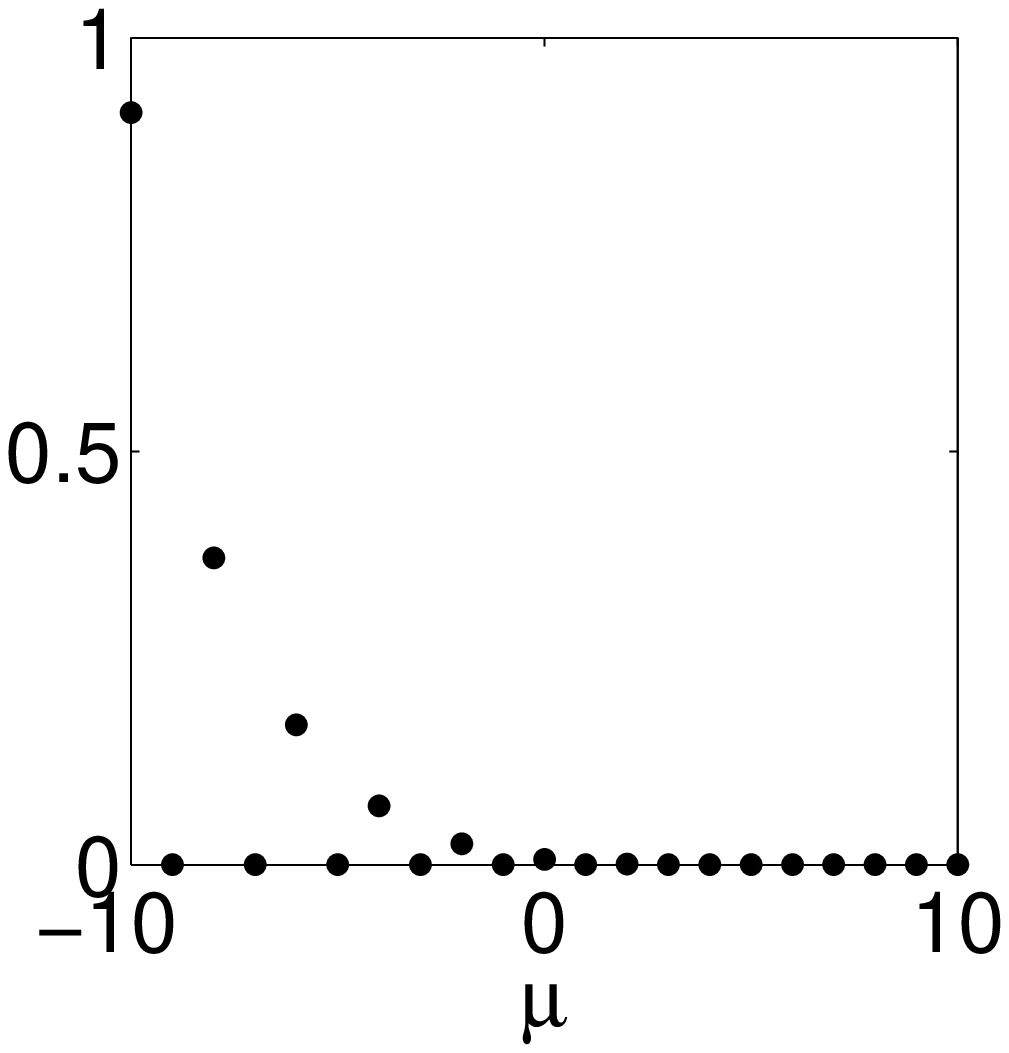}
 \end{minipage}
\begin{minipage}{0.182\textwidth}
  \includegraphics[width=\textwidth, height=\textwidth ]{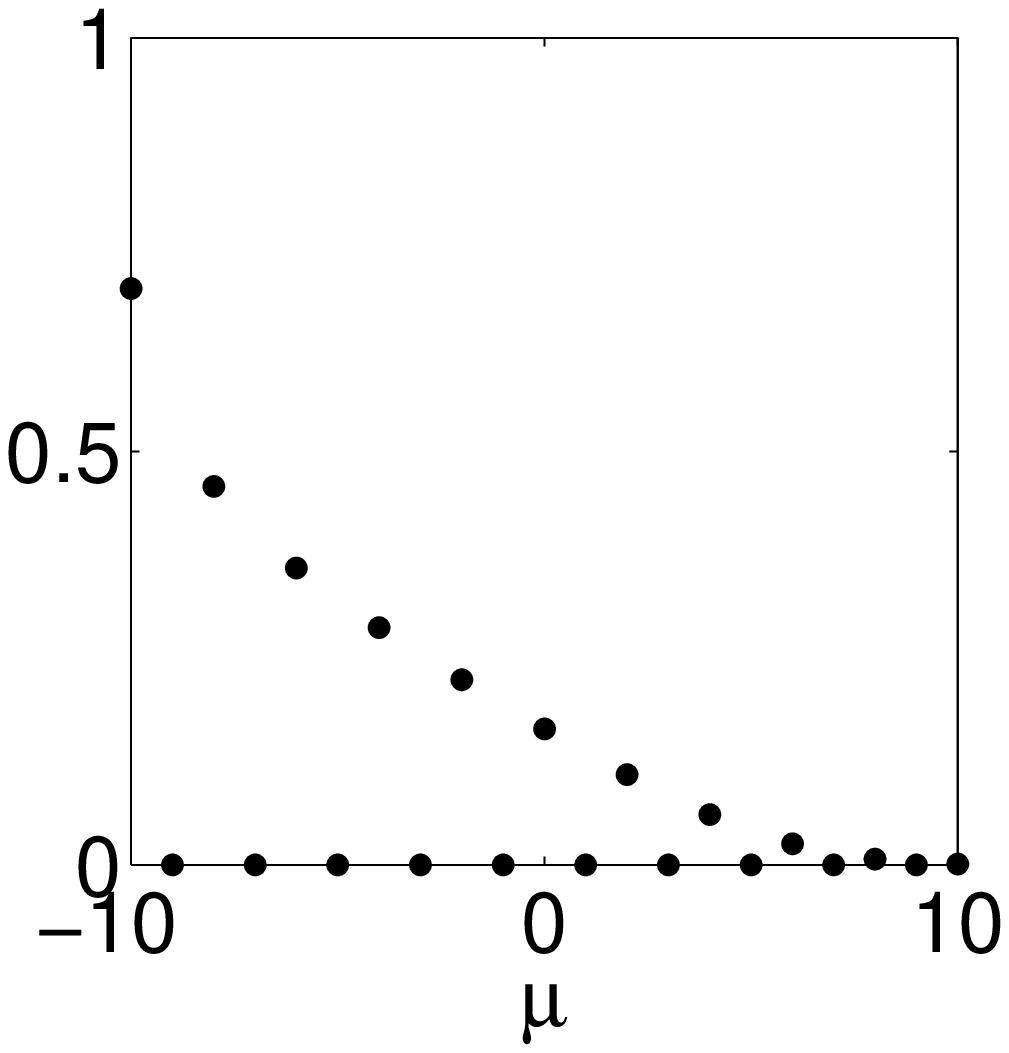}
 \end{minipage}
\begin{minipage}{0.182\textwidth}
  \includegraphics[width=\textwidth, height=\textwidth ]{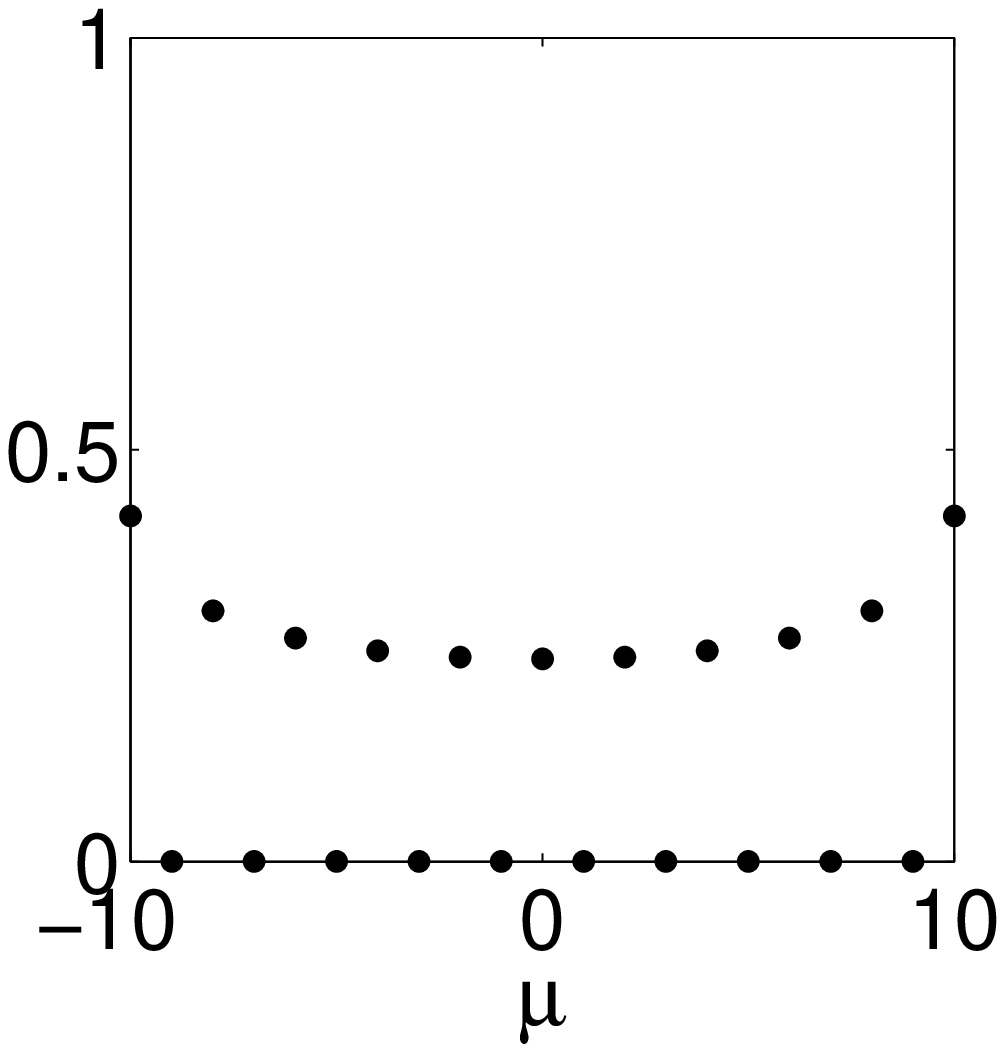}
 \end{minipage}
 \begin{minipage}{0.182\textwidth}
  \includegraphics[width=\textwidth, height=\textwidth ]{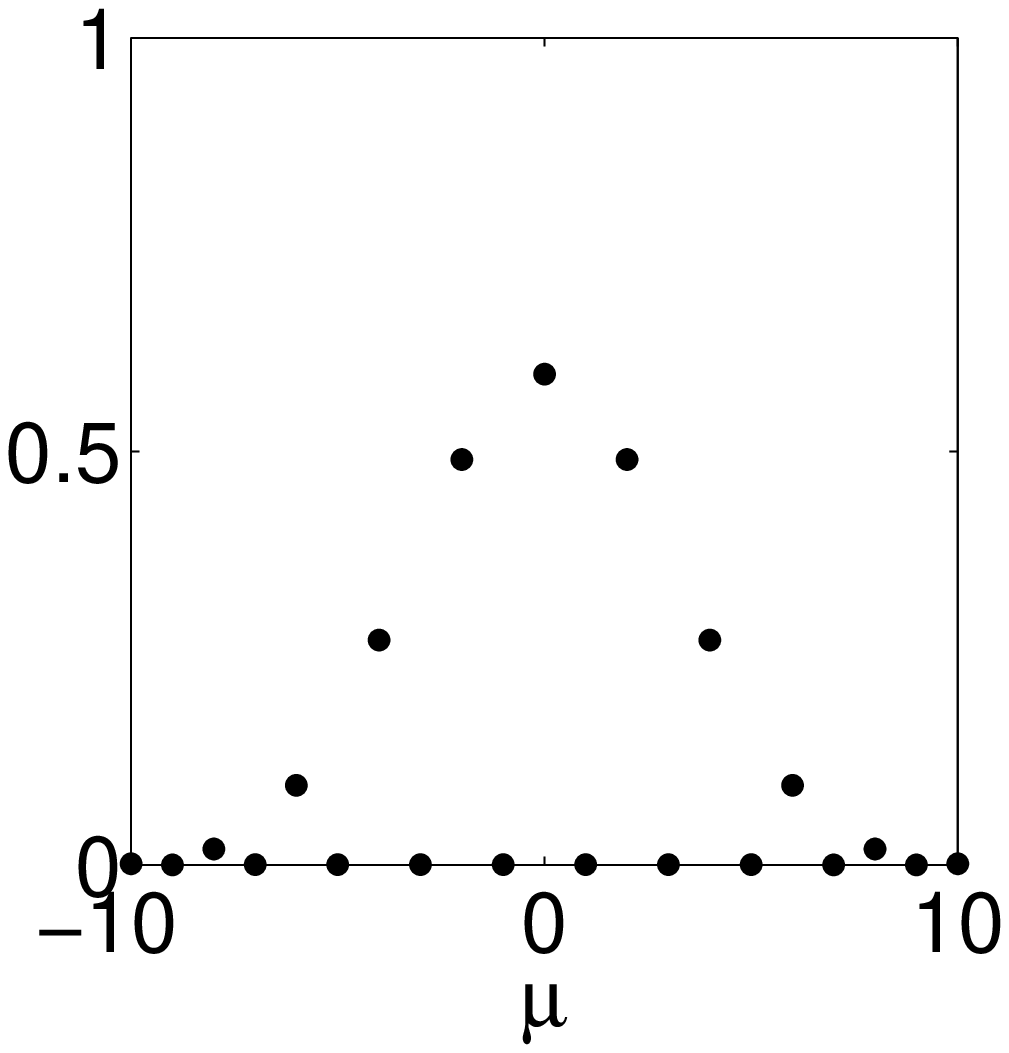}
\end{minipage}
%=====================================
% Y Row
%=====================================
\begin{minipage}{0.04\textwidth}
$_y\langle\mu|\psi\rangle$
\end{minipage}
\begin{minipage}{0.182\textwidth}
  \includegraphics[width=\textwidth, height=\textwidth ]{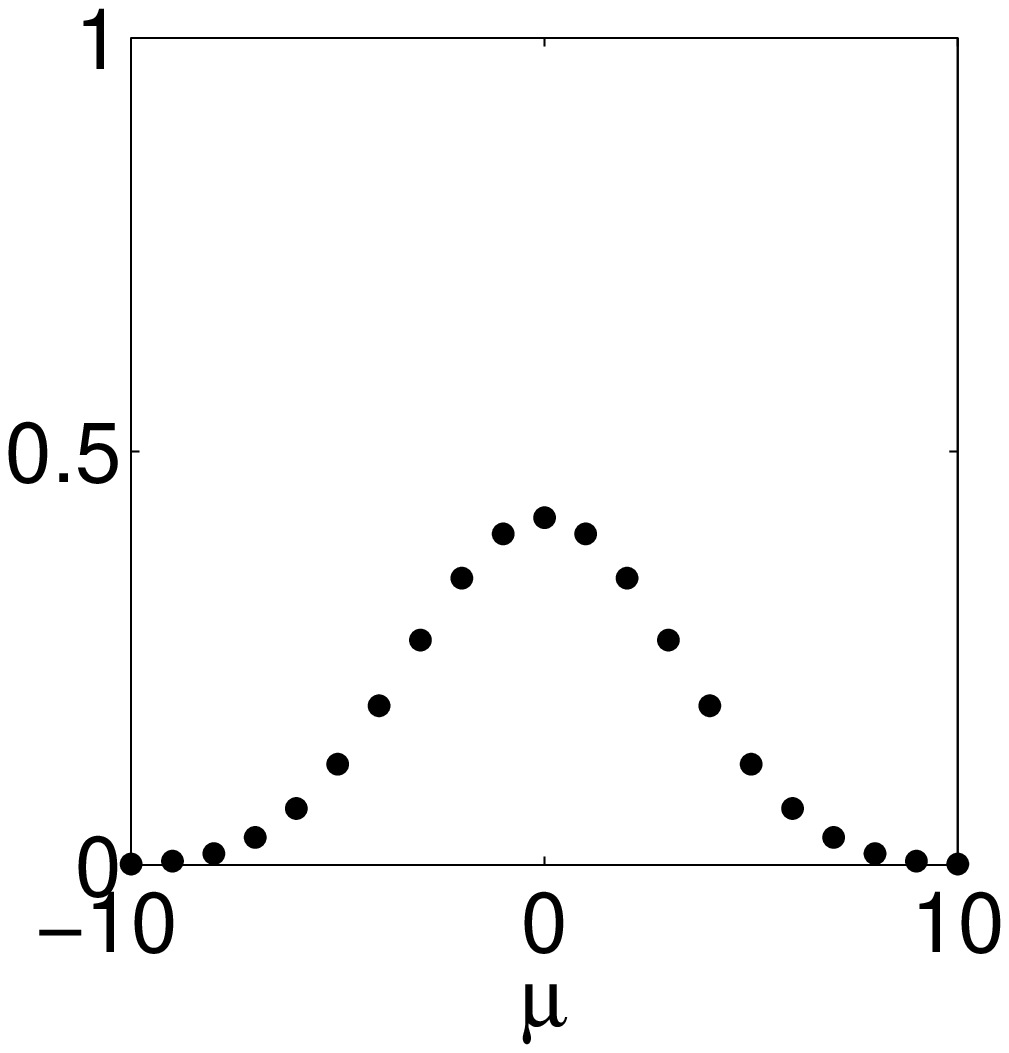}
\end{minipage}
\begin{minipage}{0.182\textwidth}
  \includegraphics[width=\textwidth, height=\textwidth ]{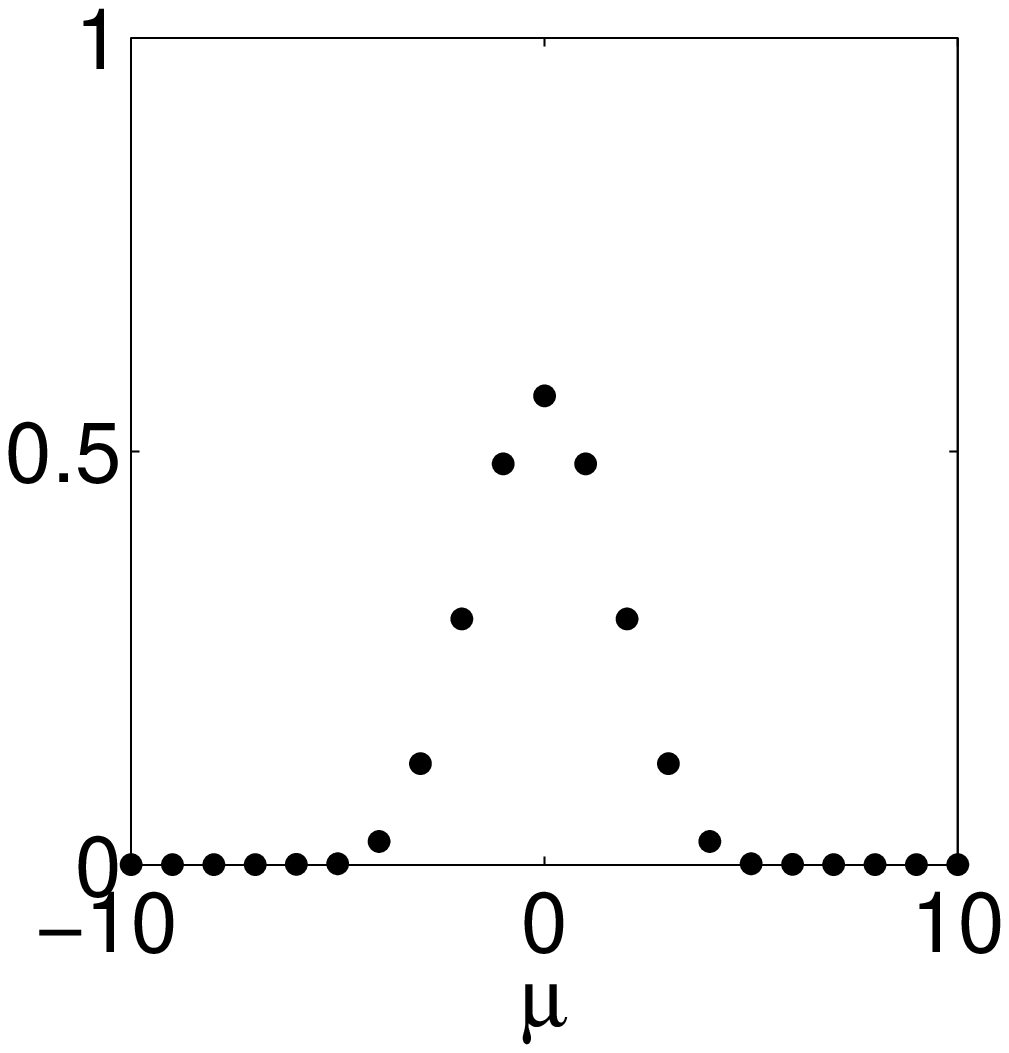}
 \end{minipage}
\begin{minipage}{0.182\textwidth}
  \includegraphics[width=\textwidth, height=\textwidth ]{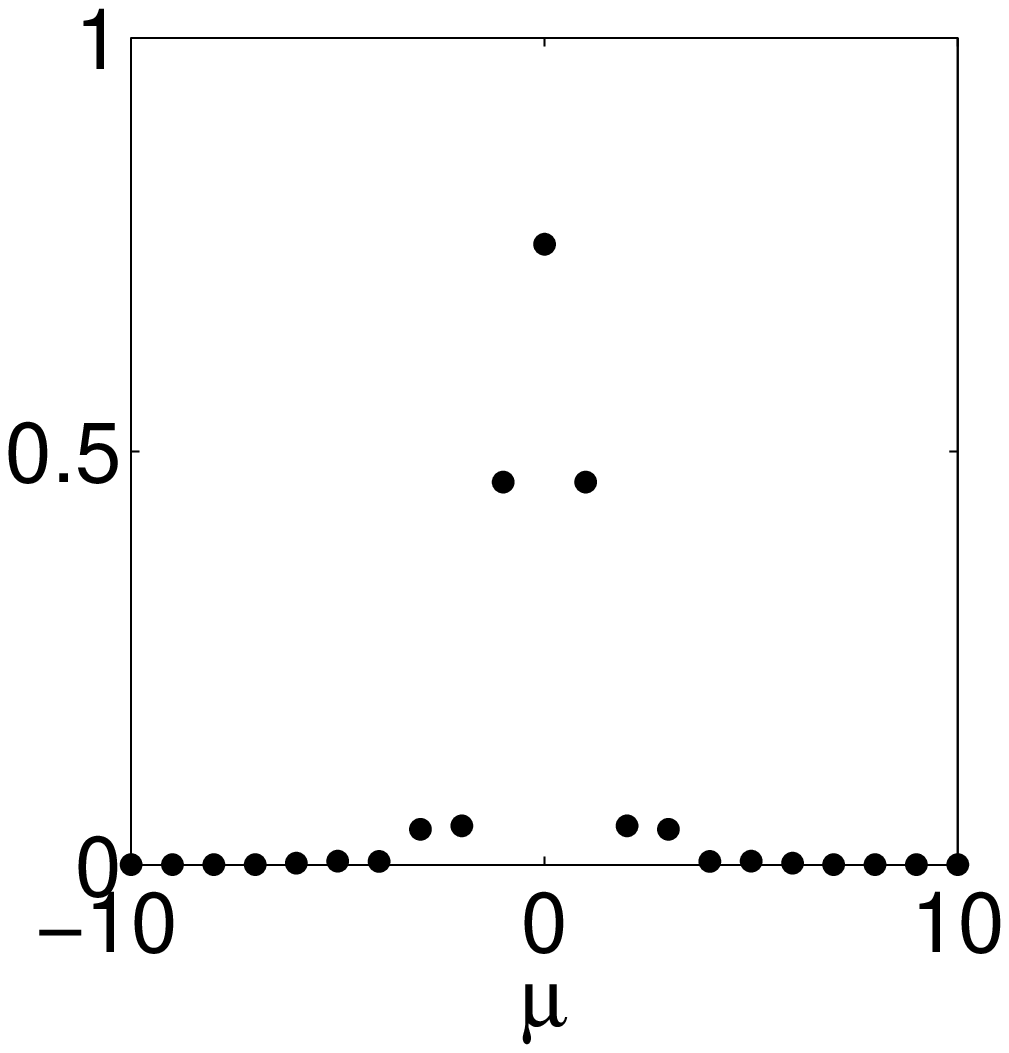}
 \end{minipage}
\begin{minipage}{0.182\textwidth}
  \includegraphics[width=\textwidth, height=\textwidth ]{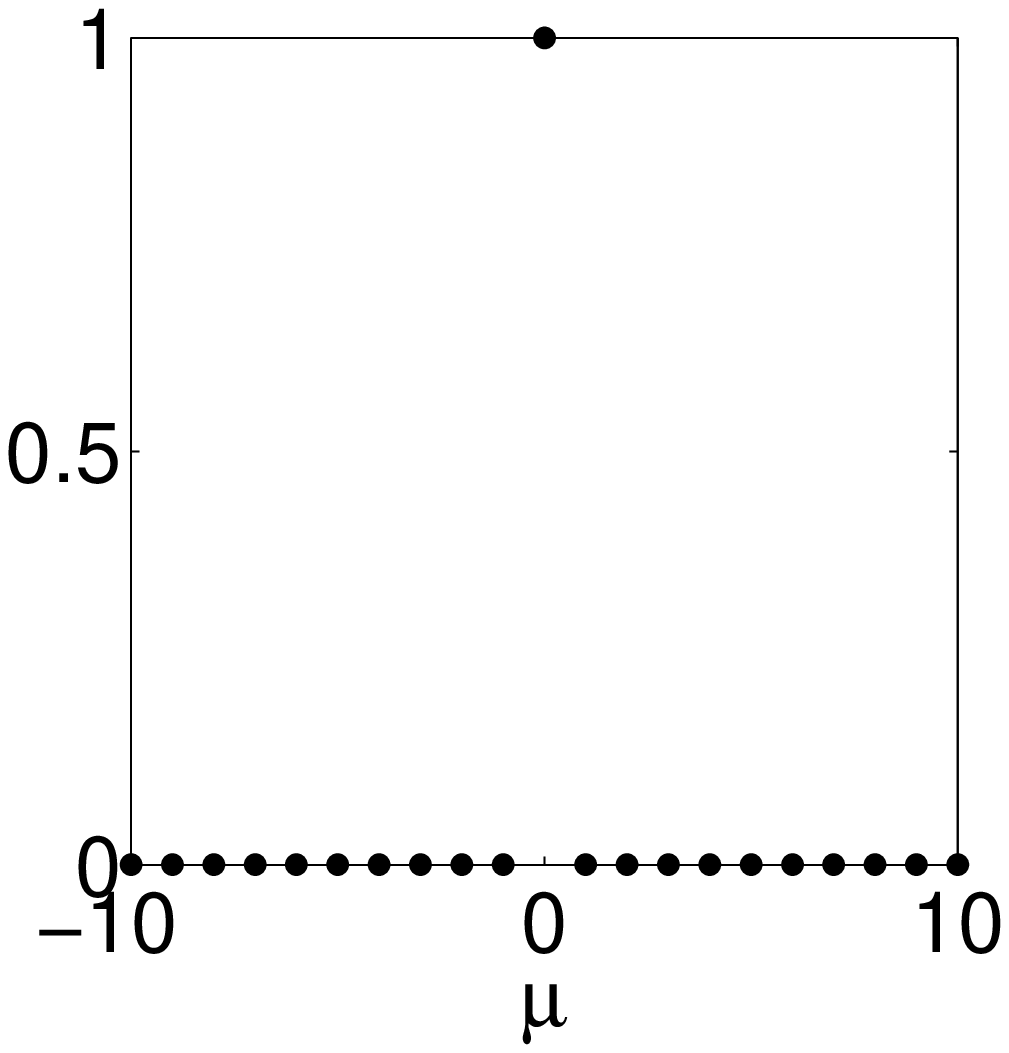}
 \end{minipage}
 \begin{minipage}{0.182\textwidth}
  \includegraphics[width=\textwidth, height=\textwidth ]{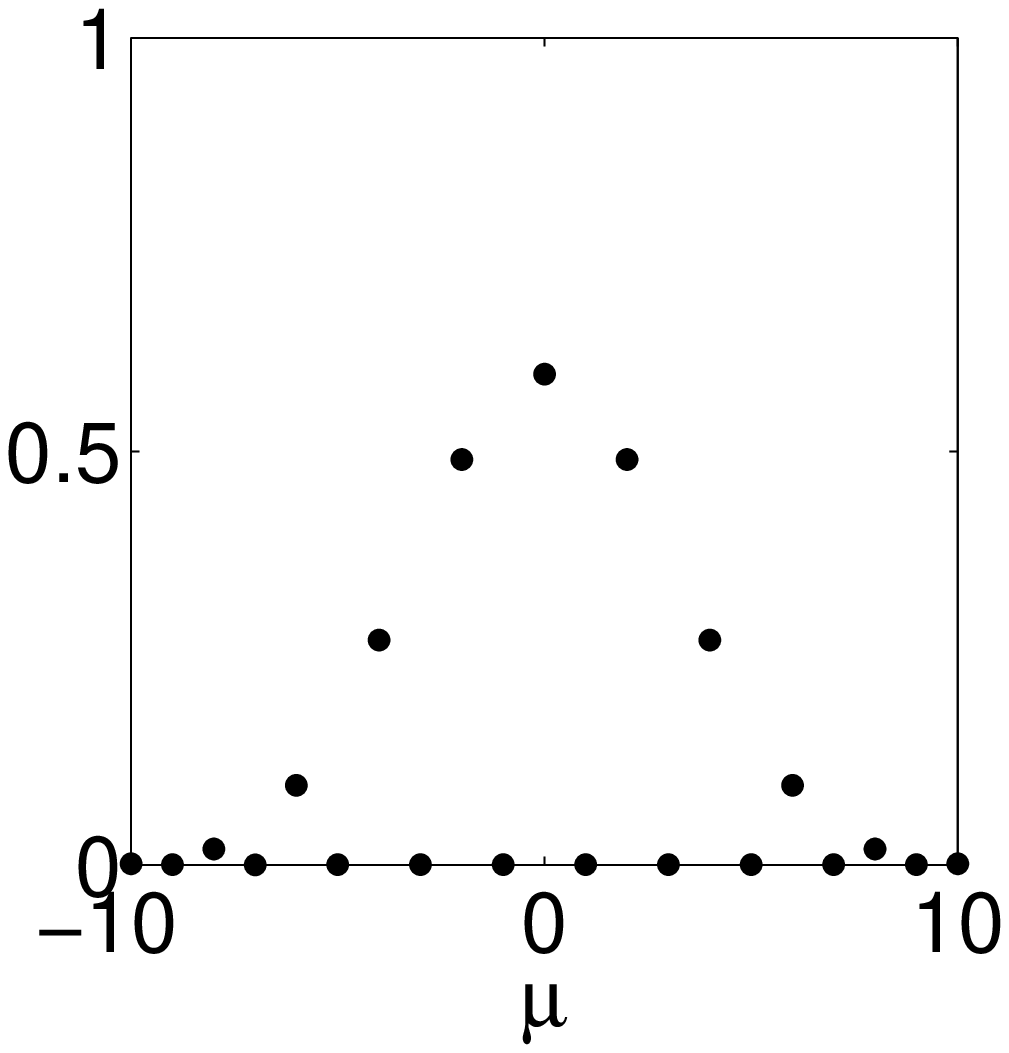}
\end{minipage}
%=====================================
% Z Row
%=====================================
\begin{minipage}{0.04\textwidth}
$_z\langle\mu |\psi\rangle$
\end{minipage}
\begin{minipage}{0.182\textwidth}
  \includegraphics[width=\textwidth, height=\textwidth ]{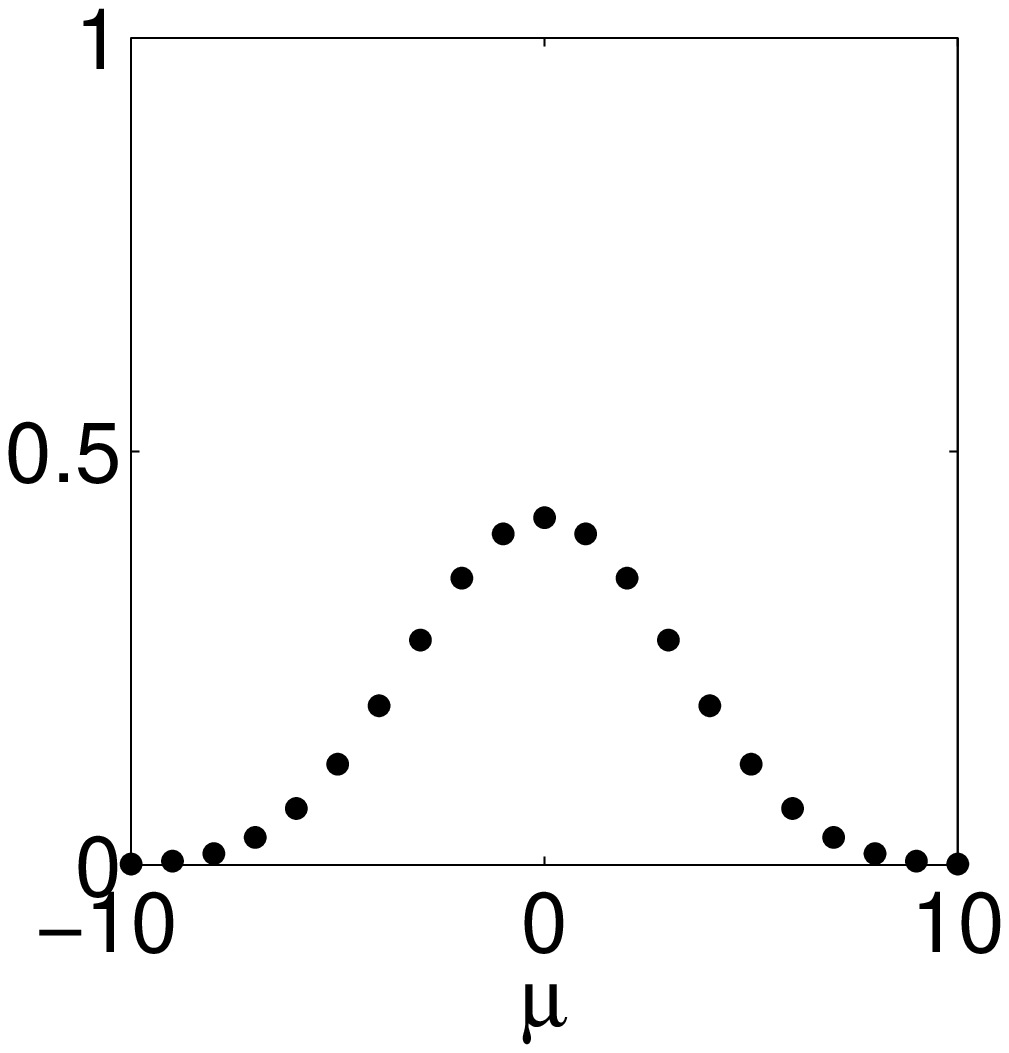}
\end{minipage}
\begin{minipage}{0.182\textwidth}
  \includegraphics[width=\textwidth, height=\textwidth ]{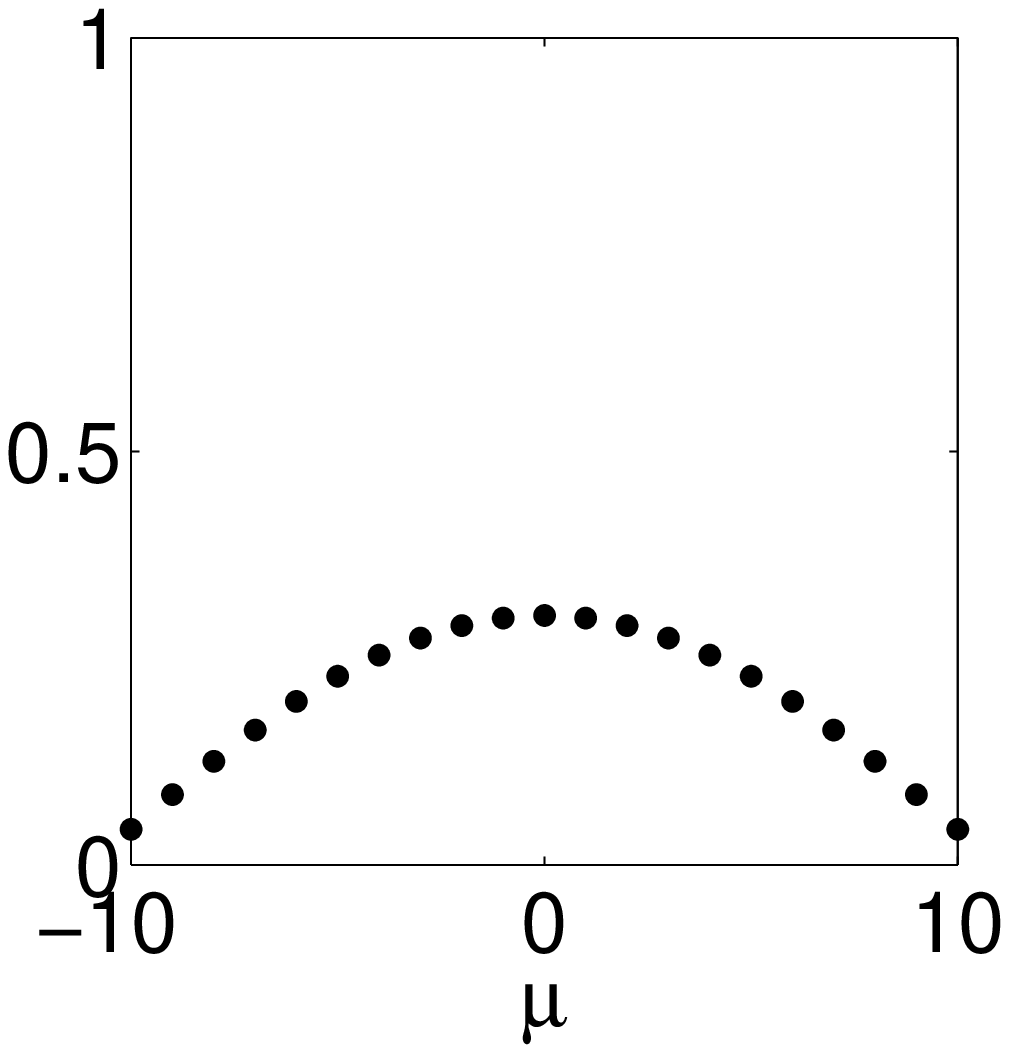}
 \end{minipage}
\begin{minipage}{0.182\textwidth}
  \includegraphics[width=\textwidth, height=\textwidth ]{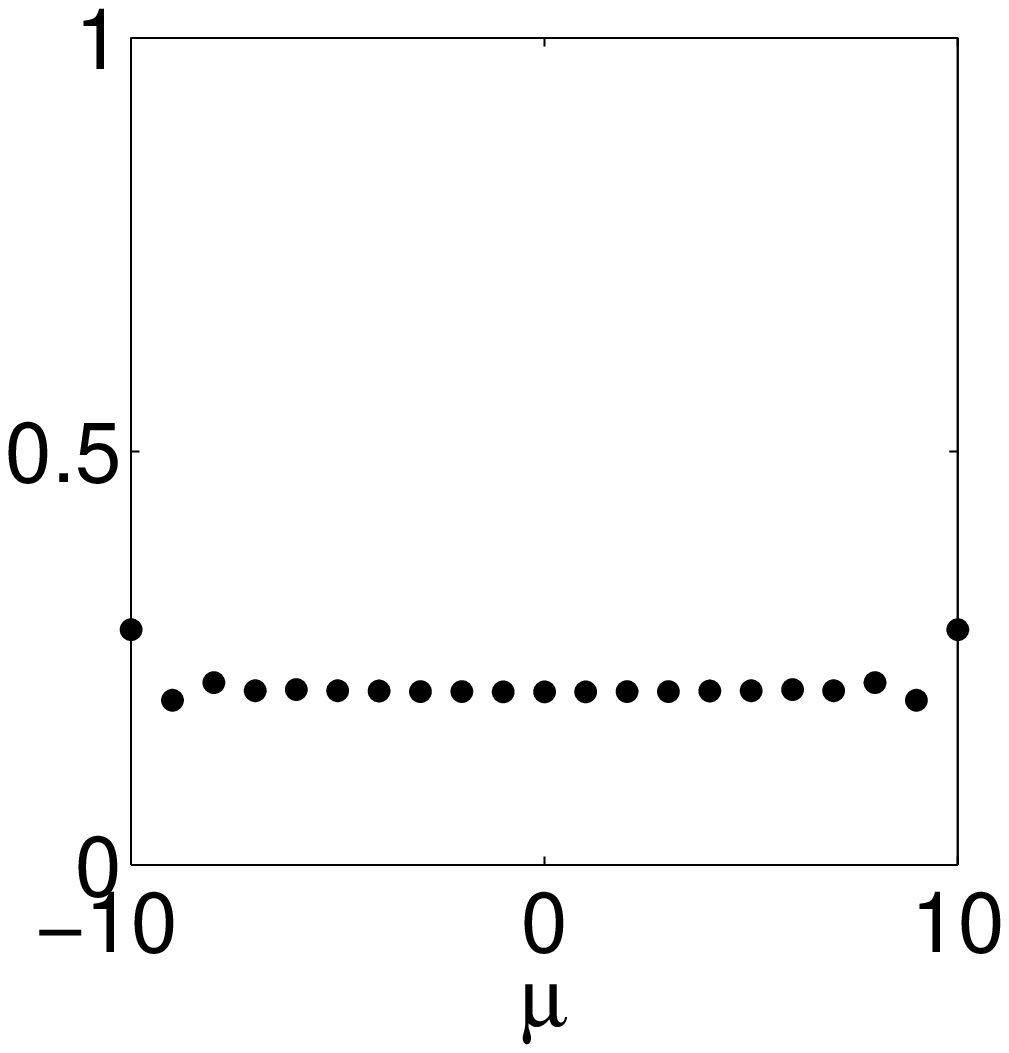}
 \end{minipage}
\begin{minipage}{0.182\textwidth}
  \includegraphics[width=\textwidth, height=\textwidth ]{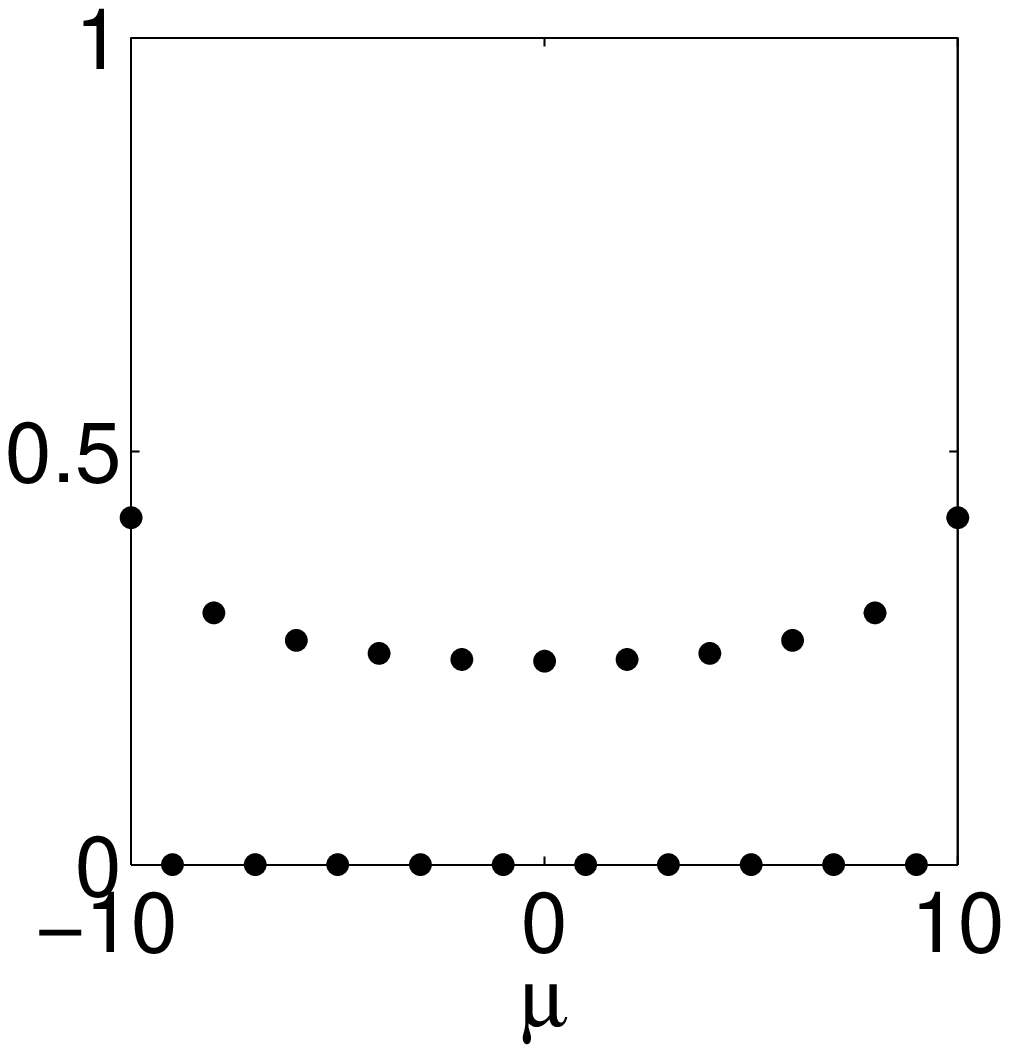}
 \end{minipage}
 \begin{minipage}{0.182\textwidth}
  \includegraphics[width=\textwidth, height=\textwidth ]{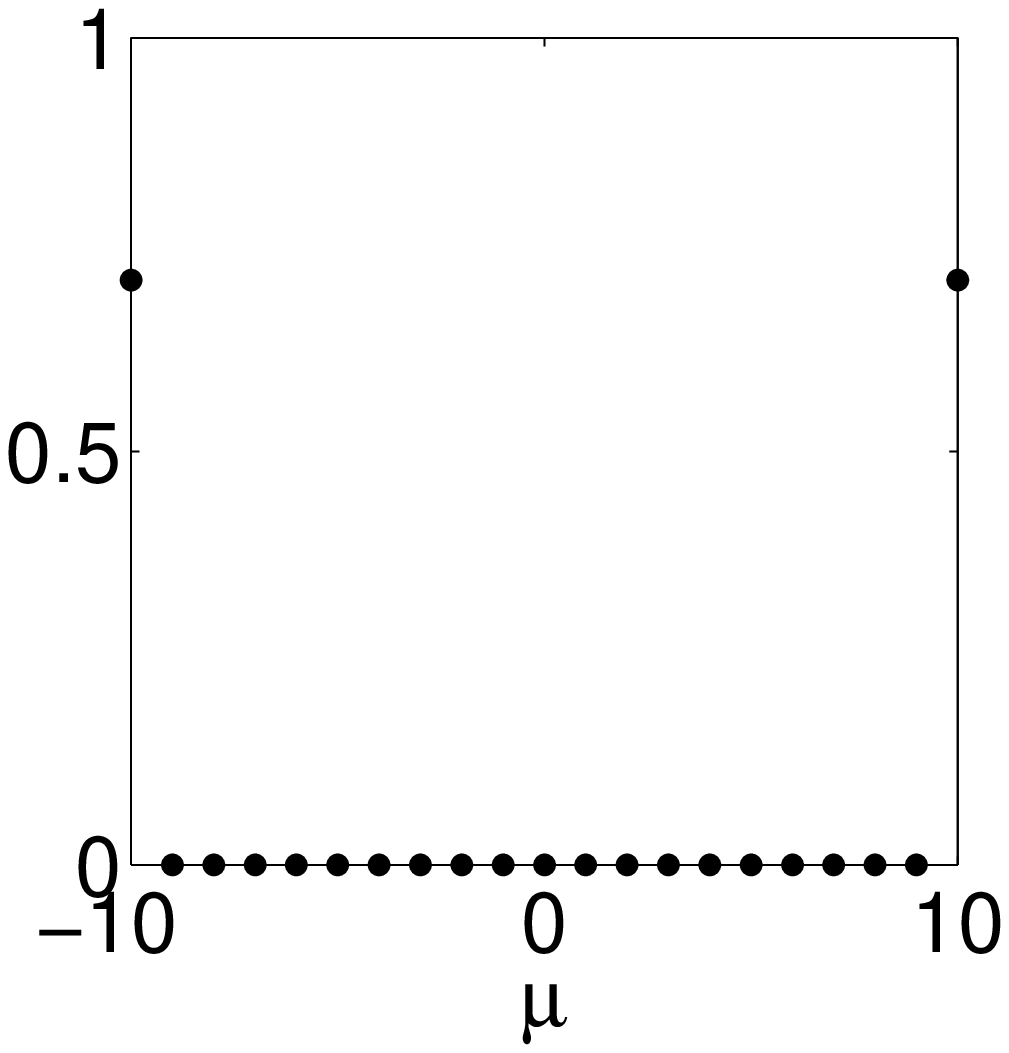}
\end{minipage}
\end{center}
\caption{In this figure the five test states (plotted for $N=20$) are in the
columns; from left to right: $\ket{\psi_{\rm coh}}$ from \erf{cohpsi},
$\ket{\psi_{\rm opt}}$ from \erf{optpsi}, $\ket{\psi_{\rm sss}}$ from
\erf{ssspsi}, $\ket{\psi_{\rm yur}}$ from \erf{yurkpsi}, and $\ket{\psi_{\rm
NOON}}$ from \erf{suppsi}. The rows contain the properties of these states that
are of interest. The first row is an equal-area plot of the Wigner function
$W(\phi,\theta)$, the second is the phase distribution $(\phi,P(\phi))$, while
the final three rows are the angular momentum coefficients  $\langle J,\mu
|\psi\rangle$  in the $x$,$y$ and $z$ directions respectively.   \label{fig0}}
\end{figure}
\end{widetext}

The final property we discuss is actually the first one (top row) plotted
in Fig.~\ref{fig0}: the Wigner function. This is a complete representation
of the quantum state, (like the coefficients in a particular direction
for a pure state). It has the advantage of showing all the properties
of the different states in a dramatic and graphical way.

The spin Wigner function
$W(\theta,\phi)$ is a pseudoprobability distribution on the Bloch sphere,
with $\theta$ and $\phi$ the usual Euler angles. For
 spin systems it is defined in Ref.~\cite{varillybondia} as
\beq\label{wignerfn}
W(\theta,\phi) = {\rm Tr}[\rho\Delta(\theta,\phi)]\\
\eeq
Here $\theta \in [-\pi/2, \pi/2]$ and $\phi \in [0,
2\pi)$ and
\begin{equation}\label{deltaeqn}
  \Delta(\theta,\phi) = \sum_{\mu,\mu'=-J}^{J}Z_{\mu,\mu'}(\theta,\phi)|J,\mu\rangle_z\langle
  J,\mu'|,
\end{equation}
Here
\begin{eqnarray}\label{zrs}
  Z_{r,s}(\theta,\phi)&=&\frac{\sqrt{4\pi}}{2j+1}\sum_{l=0}^{2j}\sqrt{2l+1}\langle j,l,r,(s-r)|j,s\rangle \nl{\times}
  Y_{l,s-r}(\theta,\phi),
\end{eqnarray}
where $\langle j,l,r,(s-r)|j,s\rangle$,  are Clebsch-Gordan coefficients and
$Y_{l,s-r}(\theta,\phi)$ is the usual spherical harmonic function.

We plot the Wigner function using the equal-area projection (described by Euclidean co-ordinates
$\phi$ and $\cos\theta$).
The original Wigner function \cite{Wig32} $W(x,p)$ for position and momentum has the property that the marginal distribution for $x$ is the true position distribution $P(x)$, and likewise for $p$.
It might be thought that the marginal distribution $\int_{-1}^1 d(\cos\theta) W(\phi,\theta)$
should equal the phase distribution $P(\phi)$. %HERE
Unfortunately this is not the case, as it follows from
Ref.~\cite{varillybondia} that the phase distribution for state $\ket{\psi}$ is actually
\beq
P_\psi(\phi) \propto \int_{-1}^1 d(\cos\theta) d\varphi W_\psi(\varphi,\theta) W_\phi(\varphi,\theta)
\eeq
where $W_\phi(\varphi,\theta)$ is the Wigner function for a phase state,
\erf{phasestates}. Because of the finiteness of the Hilbert space (unlike the $x$-$p$ case),
$W_\phi(\varphi,\theta)$ is not proportional to $\delta(\varphi-\phi)$. However, for $J$ large
it becomes very narrow in $\varphi-\phi$ and so the marginal distribution does
approximate the true phase distribution.

In the Wigner representation, the trends are quite clear. The coherent
state is approximately Gaussian with standard deviation of order $1/N$.
In the optimal state, the distribution is squeezed in $\phi$, and so antisqueezed in
$z \propto \cos\theta$. In the optimal 2ACT   spin-squeezed state, the phase squeezing has become so pronounced that the antisqueezing has produced a significant disrtribution at $\cos\theta = \pm 1$. This indicates that there is a part of the state in a superposition of $\ket{J,-J}_z$ and $\ket{J,J}_z$. This leads to the small ripples in the Wigner function  which alternate between positive and negative values (as a function of $\phi$), characteristic of the superposition in the conjugate variable $J_z$. This explains the oscillations seen in the wings of the $P(\phi)$ distribution. In the Yurke state the squeezing has become so large that the state has completely wrapped around the Bloch sphere: the distribution is equally weighted at $\phi = 0$ and $\phi=\pi$, and also at $\cos\theta = \pm 1$.  The ripples are now very pronounced, but not equal in size.  Finally, in the NOON state the distribution is confined to $\cos\theta = \pm 1$. The ripples are again very pronounced, and are equal in size, giving the sinusoidal shape of $P(\phi)$ for this state.

%==============================================================================
% Section V - lisdexic Discussion
%==============================================================================
\section{Discussion \label{Discussion}}
We have shown that for single shot measurement about which
there is no prior information, the spin squeezing parameter $\xi^2$
is not a good figure of merit. This motivated out introduction of a new parameter,
$\zeta^2$, which we call the phase squeezing. Like $\xi^2$, it can scale inversely
with the number of particles. We have also shown
that the optimal states (minimum $\zeta^2$) can be simply produced in an
ensemble of spins by starting with a coherent state (all spins pointing
in the same direction) under the well-known 2-axis counter-twisting (2ACT)
Hamiltonian. This state is different from the optimal state for spin squeezing
as produced by this Hamiltonian, which is also different from the
globally optimal spin-squeezed states.

There have been a variety of proposals for designing a system with a
 non linear Hamiltonian \cite{expHSSP, expKMB, Sorensen1}.
 So far, however the greatest degree of
 spin (or phase) squeezing observed has been created using quantum
 measurement and feedback \cite{GerStoMab04}, as proposed in
 Refs.~\cite{ThoManWis02a,ThoManWis02b}. As pointed out in
 Ref.~\cite{ThoManWis02b}, the feedback produces an effective
 system Hamiltonian proportional to the  2-axis counter-twisting Hamiltonian.
 Thus we expect that in the ideal limit feedback based on a
 QND spin measurement could also produce a state very close to the
 optimal phase squeezed state. We note that for $N\gg 1$
 and moderate degrees of
 squeezing ($1 > \xi^2 \gg 1/N$), the phase squeezing is
 identical to the spin squeezing. Finally, it is actually easier to produce
 optimal phase-squeezed states than optimal spin-squeezed states
 because it requires a lesser amount of squeezing.

To end, we comment on another method for creating entangled states
\cite{Dowling,Fiurasek} that has recently been implemented experimentally
\cite{Steinberg}. What was done in this experiment was to use ``mode mashing''
to create a (postselected) NOON state for photons. Specifically, the two
polarization modes of each photon play the role of the two spin states. The
procedure in this experiment can be very simply modified to produce other
entangled states such as Yurke states and optimal phase-squeezed states. At
present experiments are limited to $N=3$ photons, for which the advantages of
phase squeezed states are minimal. However, it should be possible in the
relatively near future to produce a phase squeezed state with $N>5$. Using
adaptive measurement techniques \cite{BerryWiseman1,BerryWisemanBreslin} it
would then be possible to perform single shot optical phase estimation
substantially better than the standard quantum limit.

\acknowledgments{ We wish to thank A. Steinberg and D. Pegg for discussions.
This work was supported by the Australian Research Council and the State of
Queensland.}
\newpage
\newpage

%==============================================================================
% The Bibliography
%==============================================================================

\bibliography{phasepaper_howard_v2}

\end{document}